\RequirePackage{ifpdf}
\documentclass[hyper]{JHEP3}

\pdfoutput=1
\usepackage{epsfig}
\usepackage{cite}
\usepackage{graphicx}
\usepackage{subfigure}
\usepackage{inputenc}
\usepackage{amsmath}
\usepackage{comment}

\newcommand{\beq}{\begin{equation}}
\newcommand{\eeq}{\end{equation}}
\newcommand{\beqn}{\begin{eqnarray}}
\newcommand{\eeqn}{\end{eqnarray}}
\def\df{{\rm d}}

\newcommand{\eq}[1]{Eq.~\eqref{eq:#1}}
\newcommand{\eqs}[2]{Eqs.~\eqref{eq:#1} and \eqref{eq:#2}}
\def\a{\alpha}

\def\nn{\nonumber\\}
\def\pd{\partial}

\title{Polarization Effects in Standard Model Parton Distributions at Very High Energies}

\author{Christian W.~Bauer$^{a}$ and Bryan R.~Webber$^b$\\
   $^a$Ernest Orlando Lawrence Berkeley National Laboratory, University of California, Berkeley, CA 94720, USA\\
   $^b$University of Cambridge, Cavendish Laboratory, J.J.\ Thomson Avenue, Cambridge, UK\\
        E-mail: \email {cwbauer@lbl.gov}, \email{webber@hep.phy.cam.ac.uk}
        }

\received{\today}               
\accepted{\today}               

\preprint{Cavendish-HEP-18/13}

\abstract{We update the earlier work
  of Refs.~\cite{Bauer:2017isx,Bauer:2017bnh} on parton distribution
  functions in the full Standard Model to
  include gauge boson polarization, non-zero input electroweak boson
  PDFs and next-to-leading-order resummation of large logarithms.}


\begin{document} 

\section{Introduction}
\label{sec:intro}
Refs.~\cite{Bauer:2017isx,Bauer:2017bnh} presented results on
the parton distribution functions for all the Standard Model fermions and
bosons up to energy scales $q$ far above the electroweak scale $\sim
m_W$. Those results were obtained in the so-called double-logarithmic
approximation (DLA), where terms of the form $\a^n\ln^{2n}(q/m_W)$ were
resummed but subleading logarithms were not all under control.  
More precisely, given that the Sudakov factors for PDF evolution
have the general form
\beq\label{eq:SudGeneral}
\Delta(q)=\exp\left[L\,g_1(\a L)+g_2(\a L)+\a \,g_3(\a L)+\ldots\right],
\eeq
where $L=\ln(q/m_W)$,
the DLA corresponds to the first term in a perturbative expansion of
$g_1$.  This is sufficient if the size of the log satisfies $\a L^2 \sim
1$ but $\a L \ll 1$. The full functions $g_i$ determine the logarithmic terms necessary 
in the expansion when the size of the log is such that $\a L\sim
1$. In this case, the function $g_1$ sums all leading logarithms (LL), $g_2$ sums next-to-leading
logs (NLL), and so on.  In the present paper, following on from our 
recent work on fragmentation functions~\cite{Bauer:2018xag}, we
upgrade the results of \cite{Bauer:2017isx,Bauer:2017bnh} on PDFs to NLL
precision by a suitable choice of scale of the running couplings in
the DGLAP equations.  

A second aspect of PDF evolution in the full SM, not treated
in~\cite{Bauer:2017isx,Bauer:2017bnh}, is the generation of gauge
boson polarizations, even in the unpolarized proton.  As emphasised
in~\cite{Manohar:2018kfx}, the fact that left- and right-handed
fermions evolve differently in the SM, and couple differently to
positive and negative boson helicities, means that the electroweak
bosons develop substantial polarization, and even the gluon
eventually becomes polarized. We upgrade the earlier results to include these
polarizations and show their effects on the fermion PDFs.

Finally we use recent results that compute the W and Z boson PDFs at the
electroweak scale~\cite{Fornal:2018znf} using the LUX formalism~\cite{Manohar:2016nzj,Manohar:2017eqh}. 
Rather than using a vanishing initial condition for the PDF evolution, as was done 
in~\cite{Bauer:2017isx,Bauer:2017bnh}, we use the results of~\cite{Fornal:2018znf} as 
the starting point, and show the effect of varying the precise scale at which we start the 
evolution. Using non-zero initial conditions requires
the introduction of a mixed Higgs PDF that corresponds to the difference
between the Higgs and longitudinal Z boson PDFs.

This paper is organized as follows: In Sec.~\ref{sec:DGLAP_polarization} 
we present the DGLAP evolution equations used in this paper, including 
polarization effects. We also discuss how to achieve next-to-leading 
logarithmic accuracy  in the collinear evolution. In Sec.~\ref{sec:implement} 
we discuss the details of our implementation, emphasizing the inclusion of 
non-zero initial conditions for the massive electroweak gauge bosons. Our 
results are presented in Sec.~\ref{sec:results} and our conclusions in 
Sec.~\ref{sec:conc}. 

\section{The evolution of SM parton distributions with polarization}
\label{sec:DGLAP_polarization}
The general form of the evolution equations is identical to the result presented
in Ref.~\cite{Bauer:2017isx}, which we repeat here for completeness:
\beqn
\label{eq:genevol}
q\frac{\pd}{\pd q} f_i(x, q) &=& \sum_I  \frac{\alpha_{I}(q)}{\pi} \left[  P^V_{i,I}(q) \, f_i(x, q) +  \sum_j  C_{ij,I} 
\int_x^{z_{\rm max}^{ij,I}(q)} \!\!\! \df z \, P^R_{ij, I}(z) f_j(x/z, q) \right]
\,.
\eeqn
Here, $i$ denotes the particle considered (specified by the type and helicity), and the sum over $I$ goes over the different interactions in the
Standard Model, which are $I = 1, 2, 3$ for the pure ${\rm U}(1)$, ${\rm SU}(2)$ and ${\rm SU}(3)$ gauge interactions, $I =Y$ for Yukawa interactions, and 
$I = M$ for the mixed interaction proportional to
\beq
\alpha_M(q) = \sqrt{\alpha_1(q)\, \alpha_2(q)}
\,.\eeq
The first contribution, proportional to $P^V_{i,I}$, denotes the virtual contribution to the PDF evolution (the disappearance
of a flavor $i$), while the second contribution is the real contribution (the appearance of flavor $i$ due to the 
splitting of a flavor $j$). The maximum value of $z$ in the integration of the real contribution depends on the 
type of splitting and interaction, and is given by
\beq
\label{eq:zmax}
z_{\rm max}^{ij,I}(q) = \Big\{
\begin{array}{ll}
1 - \frac{m_V}{q} & {\rm for}\, I = 1, 2, \,{\rm and}\, i, j \notin V \,{\rm or}\, i, j \in V
\\
1 & {\rm otherwise}
\end{array}
\,.\eeq
Having a value of $z_{\rm max} \neq 1$ amounts to applying an infrared cutoff $m_V$, of the order of the
electroweak scale, when a $B$ or $W$ boson is emitted.  This regulates
the divergence of the splitting function for those emissions as $z\to
1$.  Such a cutoff is mandatory for $I=2$ because there are PDF
contributions that are SU(2) non-singlets.  We include the same cutoff
for $I=1$, since the $B$ and $W_3$ are mixed in the physical $Z$ and
$\gamma$ states.  The evolution equations for SU(3) are regular in the
absence of a cutoff, as hadron PDFs are color singlets. 

In the rest of this section we focus on 
the modifications necessary to take into account gauge boson
polarization, non-zero electroweak input PDFs
and next-to-leading logarithmic terms.

\subsection{Polarized splitting functions}
The particles of the Standard Model we need to consider are the  fermions with left-
and right-handed chirality, denoted by $f_{L,R}$, the helicity $\pm 1$
gauge bosons, denoted by $V_\pm$, as well as spin 0 Higgs bosons,
denoted by $H$. 

Denoting the three gauge interactions of the Standard Model
collectively by $I =G$, the splitting functions involving gauge bosons are given by
\beqn
P^R_{f_Lf_L,G}(z) &=& P^R_{f_Rf_R,G}(z) =\frac 2{1-z}-(1+z) \,, \\
P^R_{V_+f_L,G}(z) &=& P^R_{V_-f_R,G}(z)=\frac{(1-z)^2}z\,,\\
P^R_{V_-f_L,G}(z) &=& P^R_{V_+f_R,G}(z)=\frac 1z\,,\\
P^R_{f_LV_+,G}(z) &=& P^R_{f_RV_-,G}(z) = \frac 12 (1-z)^2\,,\\
P^R_{f_LV_-,G}(z) &=& P^R_{f_RV_+,G}(z) = \frac 12 z^2\,,\\
P^R_{V_+V_+,G}(z) &=& P^R_{V_-V_-,G}(z) = \frac 2{1-z}+\frac 1z -1 -z(1+z)\,,\\
P^R_{V_+V_-,G}(z) &=& P^R_{V_-V_+,G}(z) = \frac{(1-z)^3}z\,,\\
P^R_{HH,G}(z) &=& \frac 2{1-z}-2\,,\\
P^R_{V_\pm H,G}(z) &=& \frac 1z -1\,,\\
P^R_{HV_\pm,G}(z) &=& \frac 12 z(1-z)\,.
\eeqn
The factor of $1/2$ in $P^R_{fV}$ has to be included since we are
considering fermions with definite chirality.
For splitting to and from antifermions we have, from CP invariance,
\beqn
&&P^R_{\bar f_LV_+,G}(z) = P^R_{f_LV_-,G}(z)\,,\;\;
P^R_{\bar f_RV_+,G}(z) = P^R_{f_RV_-,G}(z)\,,\\
&&P^R_{V_+\bar f_L,G}(z) = P^R_{V_-f_L,G}(z)\,,\;\;
P^R_{V_+\bar f_R,G}(z) = P^R_{V_-f_R,G}(z)\,.
\eeqn
For the Yukawa interaction ($Y$), one obtains
\beqn
P^R_{ff,Y}(z) &=& \frac{1-z}{2} \,, \\
P^R_{Hf,Y}(z) &=& P^R_{ff,Y}(1-z)\,,\\
P^R_{fH,Y}(z) &=& \frac{1}{2}\,.
\eeqn
.

\subsection{Isospin and CP basis}\label{sec:TCPbasis}
Taking into account the separate helicity states of the SM gauge
bosons $g,W^+,W^-,Z^0,\gamma$ and the mixed $Z^0\gamma$ and $HH$
states, there are 8 PDFs in addition to the 52 considered
in~\cite{Bauer:2017isx}.  Classifying all these according to the total
isospin $\mathbf T$ and $\mathrm{CP}$  as the quantum numbers, the
PDFs for each set of quantum numbers required are shown in
Table~\ref{tab:T_CPStates}. 
\begin{table}[h!]
\begin{center}
\begin{tabular}{l|l}
$\{\mathbf T, \mathrm{CP}\}$ & fields\\\hline
$\{0,  \pm\}$ & $2 n_g\times q_R\,, n_g\times \ell_R\,, n_g\times q_L\,, n_g\times \ell_L\,, g\,, W\,, B\,, H$  \\
$\{1,  \pm\}$ & $n_g\times q_L\,, n_g\times \ell_L\,, W\,,BW, H\,, HH$ \\
$\{2, \pm\}$ & $W$ \\
\end{tabular}
\end{center}
\caption{\label{tab:T_CPStates} The 60 PDFs required for the SM evolution can written in a basis with definite conserved quantum numbers. $2(5 n_g+4)$ FFs contribute to the $\{0, \pm\}$ states, $2(2 n_g+4)$ to each to the $\{1, \pm\}$ and 2 to the $\{2, \pm\}$, where $n_g = 3$ stands for number of generations. }
\end{table}

In terms of the states of definite flavor, the explicit  PDFs in this
basis are as follows.
Writing a fermion PDF with given $\{{\mathbf T},\mathrm{CP}\}$ as
$f_i^{\mathbf T\mathrm{CP}}$, the left-handed fermion PDFs are
\begin{align}
\label{eq:fLIsospin}
f^{0\pm}_{f_L} &= \frac
                 14\left[\left(f_{u_L}+f_{d_L}\right)\pm\left(f_{{\bar
                 u}_L}+f_{{\bar d}_L}\right)\right],\\
\label{eq:fLiso1}
f^{1\pm}_{f_L} &= \frac
                 14\left[\left(f_{u_L}-f_{d_L}\right)\pm\left(f_{{\bar
                 u}_L}-f_{{\bar d}_L}\right)\right],
\end{align}
where $u_L$ and $d_L$ refer to left-handed up- and down-type
fermions. Right-handed fermion PDFs are given by
\beq
\label{eq:fRIsospin}
f^{0\pm}_{f_R} = \frac 12\left(f_{f_R}\pm f_{{\bar f}_R}\right)\,.
\eeq
The SU(3) and U(1) boson PDFs have ${\mathbf T} = 0$, with the
unpolarized and helicity asymmetry combinations having ${\mathrm{CP}}=+$
and $-$, respectively:
\begin{align}
f^{0\pm}_g &= f_{g_+}\pm f_{g_-}\,,
     &f^{0\pm}_B &= f_{B_+}\pm f_{B_-}\,.
\end{align}
The SU(2) bosons can have $\{\mathbf T,{\mathrm{CP}}\} = \{0,+\},
\{1,-\}, \{2,+\}$ for the unpolarized PDFs and  $\{0,-\}, \{1,+\},
\{2,-\}$ for the asymmetries:
 \beqn
f^{0\pm}_W&=& \frac 13\left[\left(f_{W_+^+}+f_{W_+^-}+f_{W_+^3}\right)
\pm\left(f_{W_-^+}+f_{W_-^-}+f_{W_-^3}\right)\right],\\
f^{1\pm}_W &=& \frac 12\left[\left(f_{W_+^+}-f_{W_+^-}\right)
\mp\left(f_{W_-^+}-f_{W_-^-}\right)\right],\\
f^{2\pm}_W&=& \frac 16\left[\left(f_{W_+^+}+f_{W_+^-}-2f_{W_+^3}\right)
\pm\left(f_{W_-^+}+f_{W_-^-}-2f_{W_-^3}\right)\right].
\eeqn
The mixed $BW$ boson PDFs are a combination of $0^-$ and $1^-$ states, and
therefore they have the opposite CP to the corresponding $W$ boson PDFs: 
\beq
f^{1\pm}_{BW} = f_{BW_+}\pm f_{BW_-}\,.
 \eeq
The relations between the PDFs of $B, W^3$ and $BW$ in the unbroken
basis and those of $\gamma, Z$ and $Z\gamma$ in the broken basis were
given in~\cite{Bauer:2017isx}.

For the unmixed Higgs boson PDFs, one writes similarly to the fermions
 \beqn\label{eq:HIsospin}
&&f^{0\pm}_H = \frac
14\left[\left(f_{H^+}+f_{H^0}\right)\pm\left(f_{H^-}
      +f_{\bar H^0}\right)\right],\\
&&f^{1\pm}_H = \frac
14\left[\left(f_{H^+}-f_{H^0}\right)\pm\left(f_{H^-}
      -f_{\bar H^0}\right)\right].
\eeqn
In terms of these, the longitudinal W boson PDFs are
\begin{align}
f_{W^+_L} &=f_H^{0+}+f_H^{1+}+f_H^{0-}+f_H^{1-}\,,\\
f_{W^-_L} &=f_H^{0+}+f_H^{1+}-f_H^{0-}-f_H^{1-}\,.
\end{align}
In the notation of
Ref.~\cite{Ciafaloni:2005fm}, the neutral Higgs fields are
\beq
H^0 = \frac{(h-iZ_L)}{\sqrt 2}\,,\qquad  \bar H^0 = \frac{(h+iZ_L)}{\sqrt 2}\,,
\eeq
where $h$ and $Z_L$ represent the Higgs and the longitudinal $Z^0$
fields, respectively.  The corresponding PDFs are
\beqn
f_{H^0} &=& \frac 12\left[f_{h}+f_{Z_L}+i\left(f_{hZ_L}-f_{Z_Lh}\right)\right]\,,\\
f_{\bar H^0} &=& \frac 12\left[f_{h}+f_{Z_L}-i\left(f_{hZ_L}-f_{Z_Lh}\right)\right]\,,
\eeqn
and one can also define the mixed PDFs
\beqn
f_{H^0 \bar H^0} &=& \frac 12\left[f_{h}-f_{Z_L}-i\left(f_{hZ_L}+f_{Z_Lh}\right)\right]\,,\\
f_{\bar H^0 H^0} &=& \frac 12\left[f_{h}-f_{Z_L}+i\left(f_{hZ_L}+f_{Z_Lh}\right)\right]
\,.\eeqn
Both of these mixed Higgs PDF carry non-zero hypercharge, such 
that they are not produced by DGLAP evolution in the unbroken gauge
theory.  However, they can be present in the input at the electroweak scale $q_0$,
since the proton is an object in the broken theory.  They have isospin
1 and we can form the combinations with definite CP,
\beq
f^{1\pm}_{HH} = \frac 12\left(f_{H^0\bar H^0} \pm f_{\bar H^0 H^0}\right).
\eeq
Then the longitudinal Z and Higgs PDFs are given by
\begin{align}
f_{Z_L}&=f_H^{0+}-f_H^{1+}-f_{HH}^{1+}\,,\\
f_{h}  &=f_H^{0+}-f_H^{1+}+f_{HH}^{1+}\,.
\end{align}
There are also the mixed $hZ_L$ PDFs
\begin{align}
f_{hZ_L}+f_{Z_Lh} &=2if_{HH}^{1-}\,,\\
f_{hZ_L}-f_{Z_Lh} &=2i\left(f_H^{0-}-f_H^{1-}\right).
\end{align}
Assuming that the Higgs PDF is absent at the input scale $q_0$, we
have the following matching conditions at that scale:
\begin{align}
f_H^{0+} &=\frac 14\left(f_{W^+_L}+f_{W^-_L}+f_{Z_L}\right),\\
f_H^{0-} &=f_H^{1-} =\frac 14\left(f_{W^+_L}-f_{W^-_L}\right),\\
f_H^{1+} &=\frac 14\left(f_{W^+_L}+f_{W^-_L}-f_{Z_L}\right),\\
f_{HH}^{1+} &= -\frac 12 f_{Z_L}\,,\;\;\; f_{HH}^{1-} = 0\,.
\end{align}
Since the $f_{HH}^{1-}$ PDF is zero on input and does not mix with
any others, it remains zero and we do not consider it further.

\subsection{Upgrading to next-to-leading logarithmic accuracy}
As discussed in~\cite{Bauer:2018xag} for the case of fragmentation
function evolution, full LL resummation can be obtained in the DGLAP
formalism by choosing the scale of the
running SU(2) coupling appropriately. It is well known in standard QCD
resummation and parton shower algorithms, that for double
logarithmically sensitive observables the evolution should be
angular-ordered and the running coupling should be evaluated at the
transverse momentum of gauge boson
emission~\cite{Dokshitzer:1978hw,Amati:1980ch}.  This means that instead of using
$\alpha_2(q)$ as we have been doing in the DGLAP evolution, one should
use $\alpha_2(q(1-z))$.  Then since
\beq\label{eq:a2run}
\a_2(q') = \frac{\a_2(q)}{1+\beta^{(2)}_0\frac{\a_2(q)}{\pi}\ln\frac{q'}{q}}
\,,
\eeq
with $\beta^{(2)}_0=19/12$,
the ratio of these two scale choices is given by the expansion
\begin{align}
\frac{\alpha_2(q(1-z))}{\alpha_2(q)} = 1 - \frac{\alpha_2(q)}{\pi} \beta_0^{(2)} \ln (1-z) + \left[\frac{\alpha_2(q)}{\pi} \beta_0^{(2)} \ln (1-z)\right]^2 + \ldots \,.
\end{align}
Note that these logarithmic terms in $1-z$ only give rise to large
logarithms if integrated against a singular function $f(z) \sim 1 /
(1-z)$. Thus, in standard DGLAP evolution in QCD, where the soft
divergence as $z \to 1$ cancels between the virtual and real
contributions, the difference between these two scales do not lead to
logarithmic terms that need to be resummed. For the case of SU(2)
DGLAP evolution of PDFs or FFs that are not iso-singlets, however,
this cancelation does not happen, and one finds
\begin{align}
\int_0^{1-\frac{m}{q}} \df z \frac{\alpha_2(q(1-z))}{\pi} \frac{1}{1-z} = \frac{\alpha_2(q)}{\pi} L +  \frac{\alpha_2^2(q)}{\pi^2} \frac{\beta_0^{(2)}}{2} L^2 + \ldots
\,,
\end{align}
which generates the LL function $g_1(\a_2 L)$.  The full LL
resummation is therefore obtained by changing the SU(2) splitting functions
that are singular as $z \to 1$ as
\begin{align}
\label{eq:Pff2}
P^R_{ff,2}(z) & \to P^R_{ff,2}(z, q) = \frac{\alpha_2[q(1-z)]}{\alpha_2(q)}\frac{2}{1-z} - (1+z)\,,
\\
\label{eq:PVV2}
P^R_{V_+V_+,2}(z) & \to P^R_{VV,2}(z, q) = \frac{\alpha_2[q(1-z)]}{\alpha_2(q)}\frac{2}{1-z} + \frac{1}{z} - 1 - z(1+z)\,,\\
\label{eq:PHH2}
P^R_{HH,G}(z) & \to P^R_{HH,G}(z, q) = \frac{\alpha_2[q(1-z)]}{\alpha_2(q)}\frac{2}{1-z} - 2\,.
\end{align}

By making one more change one can in fact also reproduce the full NLL
resummation of the collinear evolution. The only missing term is the
2-loop cusp anomalous dimension, which can be included using the CMW
prescription~\cite{Catani:1990rr} for the coupling constant. This
amounts to changing
\beq\label{eq:a2CMW}
\a_2[q(1-z)] \to \alpha^{\rm CMW}_2[q(1-z)]
\eeq
in Eqs.~(\ref{eq:Pff2}-\ref{eq:PHH2}), where
\begin{align}
\alpha^{\rm CMW}_2[q(1-z)] \equiv \alpha_2[q(1-z)]\left[ 1 +
  \frac{\Gamma_{\rm cusp, f}^{(2)}}{\Gamma_{\rm cusp, f}^{(1)}}
  \frac{\alpha_2[q(1-z)]}{\pi} \right]\simeq  \alpha_2[k_{\rm CMW}q(1-z)]\,,
\end{align}
\beq
k_{\rm CMW}=\exp\left(-\frac 1{\beta_0^{(2)}} \frac{\Gamma_{\rm cusp,
      f}^{(2)}}{\Gamma_{\rm cusp, f}^{(1)}}\right)
      \,,
\eeq
and $\Gamma_{\rm cusp, f}^{(n)}$ and $\Gamma_{\rm cusp, a}^{(n)}$
denote the cusp anomalous dimension in the fundamental and adjoint
representations at $n$-loop order.  For $n_g$ fermion generations and
$n_H$ Higgs doublets~\cite{Chiu:2007dg}
\beq
 \frac{\Gamma_{\rm cusp, f}^{(2)}}{\Gamma_{\rm cusp, f}^{(1)}} =
\frac{\Gamma_{\rm cusp, a}^{(2)}}{\Gamma_{\rm cusp, a}^{(1)}} = 
\frac{67}{18}-\frac{\pi^2}6-\frac 59 n_g-\frac 19 n_H = \frac{35}{18}-\frac{\pi^2}6\,,
\eeq
which gives
\beq
k_{\rm CMW} = \exp\left(\frac{6\pi^2-70}{57}\right) = 0.828\,.
\eeq
The changes (\ref{eq:Pff2})-(\ref{eq:a2CMW}) have of course to be made in
both the real and virtual terms of the DGLAP evolution equations.
One can verify that this reproduces the complete NLL resummation in the collinear sector by comparing directly against the results of~\cite{Manohar:2018kfx}.

\subsection{Evolution equations for the various interactions}
In this section we give the complete DGLAP evolution equations, including the polarization of the vector bosons. Some of the equations of the unpolarized PDFs are identical to the results of~\cite{Bauer:2017isx,Bauer:2017bnh}, while others receive extra terms coming from the mixing with the polarization asymmetries of the vector bosons. The evolution equations for the polarization asymmetries are new. We present our results in the $\{{\mathbf T},{\mathrm{CP}}\}$ basis.

We define
\beq\label{eq:convol}
P^R_{ij,I}\otimes f_j = \int_x^{z_{\rm max}^{ij,I}(q)} \!\!\! \df z \,
P^R_{ij, I}(z) f_j(x/z, q)\,.
\eeq

For splittings involving gauge bosons, we define
\beqn
P^R_{VV,I} \otimes f_i&\equiv &\left(P^R_{V_+V_+, I}+P^R_{V_+V_-,
    I}\right)\otimes f_i\,,\\
P^R_{Vf,I} \otimes f_i&\equiv &\left(P^R_{V_+f_L, I}+P^R_{V_-f_L,
    I}\right)\otimes f_i\,,\\
P^R_{fV,I} \otimes f_i&\equiv &\left(P^R_{f_LV_+, I}+P^R_{f_LV_-,
    I}\right)\otimes f_i\,.
\eeqn
The `+'-prescription is
\beq
\label{eq:Pplusdef}
P^+_{ii, I} \otimes f_i \equiv P^R_{ii, I}\otimes f_i
+\frac{P^V_{i,I}}{C_{i,I}}f_i\,,
\eeq
where $C_{i,I}$ is the coefficient in the corresponding Sudakov factor:
\beqn
\Delta_{i,I}(q)&=&\exp\left[ \int_{q_0}^q \frac{\df q'}{q'}
  \frac{\alpha_I(q')}{\pi} P^V_{i,I}(q') \right]\nn
&=&\exp\left[ -C_{i,I}\int_{q_0}^q \frac{\df q'}{q'}
  \frac{\alpha_I(q')}{\pi} \int_0^{z_{\rm max}^{ii,I}(q)} \!\!\!
  z\,\df z\,P^R_{ii,I}(z)+\ldots\right]\,,
\eeqn
and $\ldots$ represents less divergent terms.  For convenience we also
define the isospin suppression factors
\beq\label{eq:iso_supp}
\Delta^{(T)}_{i}(q) = \left[\Delta_{i,2}(q)\right]^{T(T+1)/(2C_{i,2})}.
\eeq

For gauge bosons we also need the helicity asymmetry splitting functions:
\beqn
P^A_{VV,I} \otimes f_i&\equiv &\left(P^R_{V_+V_+, I}-P^R_{V_+V_-,
    I}\right)\otimes f_i  +\frac{P^V_{V,I}}{C_{V,I}} f_i\,,\\
P^A_{Vf,I} \otimes f_i&\equiv &\left(P^R_{V_+f_L, I}-P^R_{V_-f_L,
    I}\right)\otimes f_i\,,\\
P^A_{fV,I} \otimes f_i&\equiv &\left(P^R_{f_LV_+, I}-P^R_{f_LV_-,
    I}\right)\otimes f_i\,,
\eeqn
where the definition of $P^A_{VV,I}$ includes the plus-distribution and
\beqn
P^R_{V_+V_+, G}(z)-P^R_{V_+V_-,G}(z) &=& \frac 2{1-z}+2-4z\,,\\
P^R_{V_+f_L, G}(z)-P^R_{V_-f_L,G}(z) &=& z-2\,,\\
P^R_{f_LV_+, G}(z)-P^R_{f_LV_-,G}(z) &=& \frac 12-z\,.
\eeqn
\subsubsection{$I = 3$: SU(3) interactions}
\label{subsec:General_SU3}
We start by considering the well known case of SU(3) interactions. The
relevant degrees of freedom are the gluon, as well as left and
right-handed quarks. In the  $\{{\mathbf T},{\mathrm{CP}}\}$ basis we have 
\begin{itemize}
\item $\mathbf T = 0$ and ${\mathrm{CP}} = \pm$:
\beqn
\left[ q\frac{\pd}{\pd q} f^{0+}_{q_{L,R}} \right]_3  &=& \frac{\a_3}\pi\left[C_F P^+_{ff,G} \otimes f^{0+}_{q_{L,R}} 
+T_R P^R_{fV,G}\otimes f^{0+}_g\right],\\
 \left[ q\frac{\pd}{\pd q}f^{0+}_g \right]_3  &=& \frac{\a_3}\pi\left[C_A P^+_{VV,G}\otimes f^{0+}_g+ C_F
P^R_{Vf,G}\otimes f^{0+}_{\sum_g}\right],\\
\left[ q\frac{\pd}{\pd q} f^{0-}_{q_{L,R}} \right]_3 &=&
\frac{\a_3}\pi\left[C_F  P^+_{ff,G} \otimes f^{0-}_{q_{L,R}}
\pm T_R P^A_{fV,G}\otimes f^{0-}_g\right],\\
 \left[ q\frac{\pd}{\pd q}f^{0-}_g \right]_3 &=& \frac{\a_3}\pi\left[C_A
     P^A_{VV,G}\otimes f^{0-}_g+ C_F P^A_{Vf,G}\otimes f^{0-}_{\sum_g}\right]\,. 
\eeqn
Here $C_F = 4/3$, $C_A = 3$, $T_R = 1/2$ and
\beq
f^{0\pm}_{\sum_g} = 4 \sum_{q_L} f^{0\pm}_{q_L}\pm 2 \sum_{q_R} f^{0\pm}_{q_R}
\,,\eeq
where the sums run over all left-handed quark doublets and all right-handed quarks. The factors of $4$ and $2$ are due to the different normalizations in \eqs{fLIsospin}{fRIsospin}. 

\item All other states:
\beqn
\left[ q\frac{\pd}{\pd q} f_q \right]_3  &=& \frac{\a_3}\pi C_F P^+_{ff,G} \otimes f_q
\,.\eeqn
\end{itemize}
The virtual splitting functions are
\begin{align}
\label{eq:SudakovDef3}
P^V_{q,3}(q) &= -C_F \int_0^1 \! z \, \df z \, \left[ P^R_{ff,G}(z) + P^R_{Vf,G}(z) \right]\,,
\\
P^V_{g,3}(q) &= -\int_0^1 \! z \, \df z \,  \left[ C_A \, P^R_{VV,G}(z) + 8 \, n_g \, T_R \, P^R_{fV,G}(z)\right] 
\,,
\end{align}
where we have used in the last line that there are 8 chiral quarks
plus antiquarks per generation. 

\subsubsection{$I = 1$: U(1) interactions}
\label{subsec:General_U1}
For  ${\rm U}(1)$ the relevant degrees of freedom are left- and
right-handed fermions (denoted by the subscript $f$), the
${\rm U}(1)$ gauge boson $B$ and Higgs bosons $H$.
\begin{itemize}
\item $\mathbf T = 0$ and ${\mathrm{CP}} = +$:
\beqn
\left[ q\frac{\pd}{\pd q} f^{0+}_f \right]_1 &=& \frac{\a_1}\pi Y_f^2\left[P^+_{ff,G}\otimes f^{0+}_f
+ N_f P^R_{fV,G}\otimes f^{0+}_B\right],\\
\left[ q\frac{\pd}{\pd q} f^{0+}_B \right]_1 &=& \frac{\a_1}\pi\left[P^V_{B,1} f^{0+}_B +
P^R_{Vf,G}\otimes f^{0+}_{\sum_B f} + P^R_{VH,G} \otimes f^{0+}_H \right]\,,\\
\left[ q\frac{\pd}{\pd q} f^{0+}_H \right]_1 &=& \frac{\a_1}\pi \frac{1}{4} \left[
P^+_{HH,G}\otimes f^{0+}_H + P^R_{HV,G} \otimes f^{0+}_B\right]\,,
\eeqn
where the color factor $N_f$ is equal to 3 for quarks and 1 for
leptons, and
\beq
f^{0\pm}_{\sum_B f} = 4\sum_{f_L} Y_{f_L}^2 f^{0\pm}_{f_L}\pm 2
\sum_{f_R} Y_{f_R}^2f^{0\pm}_{f_R}
\,.
\eeq

\item $\mathbf T = 0$ and ${\mathrm{CP}} = -$:
\beqn
\left[ q\frac{\pd}{\pd q} f^{0-}_{f_{L,R}} \right]_1 &=& \frac{\a_1}\pi Y_f^2\left[P^+_{ff,G}\otimes f^{0-}_{f_{L,R}}
\pm N_f P^A_{fV,G}\otimes f^{0-}_B\right],\\
\left[ q\frac{\pd}{\pd q} f^{0-}_B \right]_1 &=& \frac{\a_1}\pi\left[P^V_{B,1} f^{0-}_B +
P^A_{Vf,G}\otimes f^{0-}_{\sum_B f}\right]\,,\\
\left[ q\frac{\pd}{\pd q} f^{0-}_H \right]_1 &=& \frac{\a_1}\pi \frac{1}{4} 
P^+_{HH,G}\otimes f^{0-}_H \,.
\eeqn

\item ${\mathbf T} = 1$ and ${\mathrm{CP}} = +$:
\beqn
\left[ q\frac{\pd}{\pd q} f^{1+}_{HH} \right]_1 &=& \frac{\a_1}\pi \frac{1}{4} 
P^+_{HH,G}\otimes f^{1+}_{HH} \,,\\
\left[ q\frac{\pd}{\pd q} f^{1+}_{BW} \right]_1 &=& \frac{\a_1}\pi \frac 12
P^V_{B,1} f^{1+}_{BW}.
\eeqn

\item ${\mathbf T} = 1$ and ${\mathrm{CP}} = -$:
\beq
\left[ q\frac{\pd}{\pd q} f^{1-}_{BW} \right]_1 =\frac{\a_1}\pi \frac 12
P^V_{B,1} f^{1-}_{BW}
\,.\eeq

\item All other states:
\beqn
\left[ q\frac{\pd}{\pd q} f_f \right]_1 &=& \frac{\a_1}\pi Y_f^2P^+_{ff,G}\otimes f_f
,\\
\left[ q\frac{\pd}{\pd q} f_H \right]_1 &=& \frac{\a_1}\pi \frac{1}{4} 
P^+_{HH,G}\otimes f_H \,.
\eeqn

\end{itemize}

The virtual splitting functions are
\begin{align}
P^V_{f,1}(q) &= -Y_f^2 \left[  \int_0^{1-\frac{m_V}{q}} \! z \, \df z \, P^R_{ff,G}(z) + \int_0^1 \! z \, \df z \,  P^R_{Vf,G}(z) \right]\,,
\\
P^V_{B,1}(q) &= -n_g \left(\frac{11}{9}N_C+3 \right) \int_0^1 \! z \, \df z \,  P^R_{fV,G}(z) -  \int_0^1 \! z \, \df z \,  P^R_{HV,G}(z)\,,
\\
P^V_{H,1}(q) &= -  \frac{1}{4} \left[ \int_0^{1-\frac{m_V}{q}} \! z \, \df z \, P^R_{HH,G}(z) + \int_0^1 \! z \, \df z \,  P^R_{VH,G}(z) \right]
\,,\end{align}
where we have used in the second line that for each generation there are 4 left-handed quarks (one needs to count particles and antiparticles separately), 2 right-handed up-type quarks, 2 right-handed down-type quarks, 4 left-handed leptons and 2 right-handed electrons, and that there are a total of 4 Higgs bosons.

\subsubsection{$I = 2$: ${\rm SU}(2)$ interactions}
\label{subsec:General_SU2}
The SU(2) interactions are more complicated, since the emission of
$W^\pm$ bosons changes the flavor of the emitting particle. This,
combined with the SU(2) breaking in the input hadron PDFs, leads to
double-logarithmic scale dependence in the DGLAP evolution, rather than
only single-logarithmic dependence as in the evolution based on U(1) and SU(3). 
The double logarithms are manifest in the appearance of the isospin
suppression factors (\ref{eq:iso_supp}). The relevant degrees of freedom are left-handed fermions, SU(2) gauge bosons $W$ and Higgs bosons. 

\begin{itemize}
\item $\mathbf T = 0$ and ${\mathrm{CP}} = +$:
\beqn\label{eq:SU2f0plus}
\left[ q\frac{\pd}{\pd q} f^{0+}_{f_L} \right]_2 &=& \frac{\a_2}{\pi}\frac 34\left[
  P^+_{ff,G}\otimes f^{0+}_{f_L}+ N_f P^R_{fV,G}\otimes f^{0+}_W\right] \,,\\
\left[ q\frac{\pd}{\pd q} f^{0+}_W \right]_2 &=& \frac{\a_2}\pi\left[2 P^+_{VV,G}\otimes
  f^{0+}_W+\sum_{f_L} P^R_{Vf,G}\otimes f^{0+}_{f_L} + P^R_{VH,G}\otimes f^{0+}_H\right] \,,
  \\
\left[ q\frac{\pd}{\pd q} f^{0+}_H \right]_2 &=& \frac{\a_2}\pi
 \frac{3}{4}  \left[ P^+_{HH,G}\otimes f^{0+}_H + P^R_{HV,G} \otimes f^{0+}_W\right]\,.
\eeqn

\item $\mathbf T = 0$ and ${\mathrm{CP}} = -$:
\beqn\label{eq:SU2f0minus}
\left[ q\frac{\pd}{\pd q} f^{0-}_{f_L} \right]_2 &=& \frac{\a_2}{\pi}\frac 34\left[
  P^+_{ff,G}\otimes f^{0-}_{f_L}+ N_f P^A_{fV,G}\otimes f^{0-}_W\right] \,,\\
\left[ q\frac{\pd}{\pd q} f^{0-}_W \right]_2 &=& \frac{\a_2}\pi\left[2 P^A_{VV,G}\otimes
  f^{0-}_W+\sum_{f_L} P^A_{Vf,G}\otimes f^{0-}_{f_L}\right] \,,
  \\
\left[ q\frac{\pd}{\pd q} f^{0-}_H \right]_2 &=& \frac{\a_2}\pi
 \frac{3}{4} P^+_{HH,G}\otimes f^{0-}_H\,.
\eeqn

\item $\mathbf T = 1$ and ${\mathrm{CP}} = +$:
\beqn
\left[ \Delta_{f}^{(1)}q\frac{\pd}{\pd q} \frac{f^{1+}_{f_L}}{\Delta_{f}^{(1)}} \right]_2 &=& \frac{\a_2}{\pi}\left[-\frac 14
  P^+_{ff,G}\otimes f^{1+}_{f_L}+\frac 12 N_f P^A_{fV,G}\otimes
 f^{1+}_{W}\right] \\
\left[ \Delta_{V}^{(1)}q\frac{\pd}{\pd
    q}\frac{f^{1+}_{W}}{\Delta_{V}^{(1)}} \right]_2 &=&
\frac{\a_2}\pi\left[P^A_{VV,G}\otimes  f^{1+}_{W}+\sum_{f_L} P^A_{Vf,G}\otimes
  f^{1+}_{f_L}\right] \\
\left[\Delta_{H}^{(1)}q\frac{\pd}{\pd
    q}\frac{f^{1+}_{H}}{\Delta_{H}^{(1)}}\right]_2 &=&
\frac{\a_2}\pi\left[ -\frac{1}{4} P^+_{HH,G} \otimes f^{1+}_{H}
\right]\\
\left[\Delta_{H}^{(1)}q\frac{\pd}{\pd
    q}\frac{f^{1+}_{HH}}{\Delta_{H}^{(1)}}\right]_2 &=&
\frac{\a_2}\pi\left[ -\frac{1}{4} P^+_{HH,G} \otimes f^{1+}_{HH}
\right]\\ \label{eq:BW1+}
\left[\Delta_{V}^{(1)}q\frac{\pd}{\pd q}\frac{f^{1+}_{BW}}{\Delta_{V}^{(1)}}\right]_2 &=& 0\,.
\eeqn

\item $\mathbf T = 1$ and ${\mathrm{CP}} = -$:
\beqn
\left[ \Delta_{f}^{(1)}q\frac{\pd}{\pd q} \frac{f^{1-}_{f_L}}{\Delta_{f}^{(1)}} \right]_2 &=& \frac{\a_2}{\pi}\left[-\frac 14
  P^+_{ff,G}\otimes f^{1-}_{f_L}+\frac 12 N_f P^R_{fV,G}\otimes
 f^{1-}_{W}\right] \\
\left[ \Delta_{V}^{(1)}q\frac{\pd}{\pd
    q}\frac{f^{1-}_{W}}{\Delta_{V}^{(1)}} \right]_2 &=&
\frac{\a_2}\pi\left[P^+_{VV,G}\otimes  f^{1-}_{W}+\sum_{f_L} P^R_{Vf,G}\otimes
  f^{1-}_{f_L} + P^R_{VH,G}\otimes f^{1-}_{H}\right] \\
  \nn
\left[\Delta_{H}^{(1)}q\frac{\pd}{\pd
    q}\frac{f^{1-}_{H}}{\Delta_{H}^{(1)}}\right]_2 &=&
\frac{\a_2}\pi\left[ -\frac{1}{4} P^+_{HH,G} \otimes f^{1-}_{H} +
  \frac{1}{2} \, P^R_{HV,G} \otimes f^{1-}_{W} \right]\,\\ \label{eq:BW1-}
\left[\Delta_{V}^{(1)}q\frac{\pd}{\pd
    q}\frac{f^{1-}_{BW}}{\Delta_{V}^{(1)}}\right]_2 &=& 0\,.
\eeqn

\item $\mathbf T = 2$ and ${\mathrm{CP}} = +$:
\beqn\label{eq:SU2f2plus}
\left[ \Delta_{V}^{(2)}q\frac{\pd}{\pd q} \frac{f^{2+}_{W}}{\Delta_{V}^{(2)}}\right]_2 &=& -\frac{\a_2}\pi P^+_{VV,G}\otimes
  f^{2+}_{W}\,.
\eeqn

\item $\mathbf T = 2$ and ${\mathrm{CP}} = -$:
\beqn\label{eq:SU2f2minus}
\left[ \Delta_{V}^{(2)}q\frac{\pd}{\pd q} \frac{f^{2-}_{W}}{\Delta_{V}^{(2)}}\right]_2 &=& -\frac{\a_2}\pi P^A_{VV,G}\otimes
  f^{2-}_{W}\,.
\eeqn

\end{itemize}
where the sum in the last line is over all left-handed fermions and anti-fermions.

The virtual splitting functions are
\begin{align}
\label{eq:SU2_virtual}
P^V_{f,2}(q) &= -\frac{3}{4} \left[  \int_0^{1-\frac{m_V}{q}} \! z \, \df z \, P^R_{ff,G}(z) + \int_0^{1} \! z \, \df z \, P^R_{Vf,G}(z) \right]\,,
\\
P^V_{W,2}(q) &= - 2  \int_0^{1-\frac{m_V}{q}} \! z \, \df z \,  P^R_{VV,G}(z) - n_g(N_C+1) \int_0^{1} \! z \, \df z \,  P^R_{fV,G}(z)  - \int_0^{1} \! z \, \df z \,  P^R_{HV,G}(z)\,,
\\
P^V_{H,2}(q) &= -  \frac{3}{4} \left[ \int_0^{1-\frac{m_V}{q}} \! z \, \df z \, P^R_{HH,G}(z) + \int_0^1 \! z \, \df z \,  P^R_{VH,G}(z) \right]
\,.\end{align}

\subsubsection{$I = Y$: Yukawa interactions}
\label{subsec:General_Y}
The interaction of Higgs particles with fermions is described by the Yukawa interactions. In this work we only keep the top Yukawa coupling, setting all others to zero. 
This gives contributions to the top quark PDFs, the left-handed bottom
PDF and the Higgs PDFs:
\begin{itemize}
\item $\mathbf T = 0$ and ${\mathrm{CP}} = +$:
\beqn
\left[q\frac{\pd}{\pd q}  f^{0+}_{q^3_L} \right]_Y &=&
\frac{\a_Y}{\pi}\biggl[P^V_{q^3_L,Y} f^{0+}_{q^3_L} + 
P^R_{ff,Y} \otimes f^{0+}_{t_R}  + N_c P^R_{fH,Y} \otimes f^{0+}_{H} \biggr]\\
\left[q\frac{\pd}{\pd q}  f^{0+}_{t_R} \right]_Y &=&
\frac{\a_Y}{\pi}\,2\,\biggl[P^V_{t_R,Y} f^{0+}_{t_R} + 
P^R_{ff,Y} \otimes f^{0+}_{q^3_L}  +N_c  P^R_{fH,Y} \otimes f^{0+}_{H}\biggr]\\
\left[ q\frac{\pd}{\pd q} f^{0+}_{H} \right]_Y &=&
\frac{\a_Y}{\pi}\biggl[P^V_{H,Y} f^{0+}_{H} + 
P^R_{Hf,Y} \otimes f^{0+}_{\sum_H f}\biggr]
\,,\eeqn
where
\beq
f^{0+}_{\sum_H f} = f^{0+}_{t_R} + f^{0+}_{q_L^3} \,.
\eeq

\item $\mathbf T = 0$ and ${\mathrm{CP}} = -$:
\beqn
\left[q\frac{\pd}{\pd q}  f^{0-}_{q^3_L} \right]_Y &=&
\frac{\a_Y}{\pi}\biggl[P^V_{q^3_L,Y} f^{0-}_{q^3_L} + 
P^R_{ff,Y} \otimes f^{0-}_{t_R}  - N_c P^R_{fH,Y} \otimes f^{0-}_{H} \biggr]\\
\left[q\frac{\pd}{\pd q}  f^{0-}_{t_R} \right]_Y &=&
\frac{\a_Y}{\pi}\,2\,\biggl[P^V_{t_R,Y} f^{0-}_{t_R} + 
P^R_{ff,Y} \otimes f^{0-}_{q^3}  +  N_c P^R_{fH,Y} \otimes f^{0-}_{H}\biggr]\\
\left[ q\frac{\pd}{\pd q} f^{0-}_{H} \right]_Y &=&
\frac{\a_Y}{\pi}\biggl[P^V_{H,Y} f^{0-}_{H} + 
P^R_{Hf,Y} \otimes f^{0-}_{\sum_H f}\biggr]
\,,\eeqn
where
\beq
f^{0-}_{\sum_H f} = f^{0-}_{t_R} - f^{0-}_{q_L^3} \,.
\eeq

\item $\mathbf T = 1$ and ${\mathrm{CP}} = +$:
\beqn
\left[ q\frac{\pd}{\pd q} f^{1+}_{q^3_L} \right]_Y &=&
\frac{\a_Y}{\pi}\biggl[P^V_{q^3_L,Y} f^{1+}_{q^3_L} - N_c P_{fH,Y} \otimes f^{1+}_{H}\biggr] \\
\left[ q\frac{\pd}{\pd q} f^{1+}_{H} \right]_Y &=&
\frac{\a_Y}{\pi}\biggl[P^V_{H,Y} f^{1+}_{H}  -
P^R_{Hf} \otimes f^{1+}_{q_L^3}\biggr]
\eeqn

\item $\mathbf T = 1$ and ${\mathrm{CP}} = -$:
\beqn
\left[ q\frac{\pd}{\pd q} f^{1-}_{t_L} \right]_Y &=&
\frac{\a_Y}{\pi}\biggl[P^V_{t_L,Y} f^{1-}_{t_L} +
N_c P_{fH,Y} \otimes f^{1-}_{H}\biggr] \\
\left[ q\frac{\pd}{\pd q} f^{1-}_{H} \right]_Y &=&
\frac{\a_Y}{\pi}\biggl[P^V_{H,Y} f^{1-}_{H} + 
 P^R_{Hf,Y} \otimes f^{1-}_{q_L^3}\biggr]
\eeqn
\end{itemize}

The virtual splitting functions are
\begin{align}
P^V_{q_L^3,Y}(q) = \frac{1}{2} P^V_{t_R,Y}(q) &= -\int_0^{1} \! z \, \df z \, P^R_{ff,Y}(z) - \int_0^{1} \! z \, \df z \, P^R_{Hf,Y}(z)\,,
\\
P^V_{H,Y}(q) &= - 2 N_C  \int_0^{1} \! z \, \df z \,  P^R_{fH,Y}(z)
\,.\end{align}

\subsubsection{$I = M$: Mixed $B - W_3$ interactions}
\label{subsec:General_M}
Finally, we need to consider the evolution involving the mixed $BW$
boson PDF.  The diagonal splittings $P^R_{ii,G}$ are absent
because  there is no vector boson with both U(1) and SU(2)
interactions. For the same reason, there are no virtual contributions
associated with the mixed interaction.

\begin{itemize}
\item $\mathbf T = 1$ and ${\mathrm{CP}} = +$:
\beqn
\left[ q\frac{\pd}{\pd q} f^{1+}_{f} \right]_M &=& \frac{\a_M}\pi
\frac{Y_f}{2} N_f P^R_{fV,G}\otimes f^{1+}_{BW}\,,\\
\label{eq:mixBW1}
\left[q\frac{\pd}{\pd q} f^{1+}_{BW}\right]_M &=& \frac{\a_M}\pi
\left[
4\sum_{f_L} Y_f P^R_{Vf,G}\otimes f^{1+}_{f}  + 2 P^R_{VH,G} \otimes f^{1+}_{H}
\right]\,,\\
\left[ q\frac{\pd}{\pd q} f^{1+}_H \right]_M &=& \frac{\a_M}\pi \frac{1}{4} P^R_{HV,G} \otimes f^{1+}_{BW}\,.
\eeqn

\item $\mathbf T = 1$ and ${\mathrm{CP}} = -$:
\beqn
\left[ q\frac{\pd}{\pd q} f^{1-}_{f_L} \right]_M &=& \frac{\a_M}\pi \frac{Y_f}{2}
N_f P^A_{fV,G}\otimes f^{1-}_{BW}\,,\\
\label{eq:mixBW2}
\left[q\frac{\pd}{\pd q} f^{1-}_{BW}\right]_M &=& \frac{\a_M}\pi
4\sum_{f_L} Y_f P^A_{Vf,G}\otimes f^{1-}_{f} \,,\\
\left[ q\frac{\pd}{\pd q} f^{1-}_H \right]_M &=& 0\,.
\eeqn

\end{itemize}

As seen in Sections~\ref{subsec:General_U1} and \ref{subsec:General_SU2}, the mixed gauge
field PDF $f_{BW}$ has U(1) and SU(2) virtual interactions with
no corresponding real emission term in its evolution
equations.  It evolves double-logarithmically and is
suppressed at high scales relative to the unmixed PDFs.

\section{Implementation details}
\label{sec:implement}
Our treatment assumes that the SM PDFs at very high energies can be
obtained by smoothly matching the broken and unbroken symmetry regimes
at a matching scale $q_0\sim m_V$. As a default, we choose $q_0 = m_V=100$ GeV, 
however we will also show some results for other values of $q_0$ and
$m_V$, to assess the sensitivity to these parameters.
Our input PDFs at $q_0$ are obtained as follows: We take
the CT14qed PDF set~\cite{Schmidt:2015zda} at 10 GeV and replace the
photon PDF by that of the LUXqed set~\cite{Manohar:2016nzj}.  We do not use
the CT14qed photon because the LUXqed photon, while being consistent with
CT14qed, has much smaller uncertanties and a smoother $x$ dependence.
The LUXqed PDF set combines the  PDF4LHC15\_nnlo\_100 parton
 set~\cite{Butterworth:2015oua}  with a determination of the photon PDF
from structure function and elastic form factor fits in electron-proton scattering.
However, we do not use the LUXqed partons, because being NNLO they are not
positive-definite, which we require for our LO treatment and is satisfied by CT14qed.

We evolve this hybrid CT14-LUX PDF set from 10 GeV to $q_0$ using leading-order QCD
plus QED evolution, which incidentally generates the charged leptons. This generates the input
of the quarks, charged leptons and the photon. The input transverse and
longitudinal electroweak boson PDFs are those computed at $q_0$ by the method of
Fornal, Manohar and Waalewijn~\cite{Fornal:2018znf}\footnote{We thank
  the authors for providing these PDFs at a range of input scales.}. The top quark,
neutrino and Higgs PDFs are taken to be zero at $q_0$. 

The resulting PDFs in the broken phase are mapped onto the
unbroken basis, as discussed in Sect.~\ref{sec:TCPbasis}, and form the input to the
unbroken SM evolution upwards from $q_0$.  All PDFs that were zero at the input are generated
dynamically.

\section{Results}
\label{sec:results}
Most plots we present in this section are very similar to those that were 
already shown in~\cite{Bauer:2017isx,Bauer:2017bnh}. This is done on 
purpose, since it allows us to highlight the differences from the
results obtained without
the updates made in the present paper. Whenever possible, we show in solid lines
the results including all effects introduced in this paper (``Best''), and in dashed lines
the results without these improvements (``Old''). Note that a few small changes
in the evolution were made between~\cite{Bauer:2017isx} and~\cite{Bauer:2017bnh},
having to do with details of how the top quark threshold is included in the running
strong coupling constant. Thus in some of the plots the dashed line does not correspond
exactly to the results presented in the previous papers. 

\FIGURE[h]{
 \centering
  \includegraphics[scale=0.45]{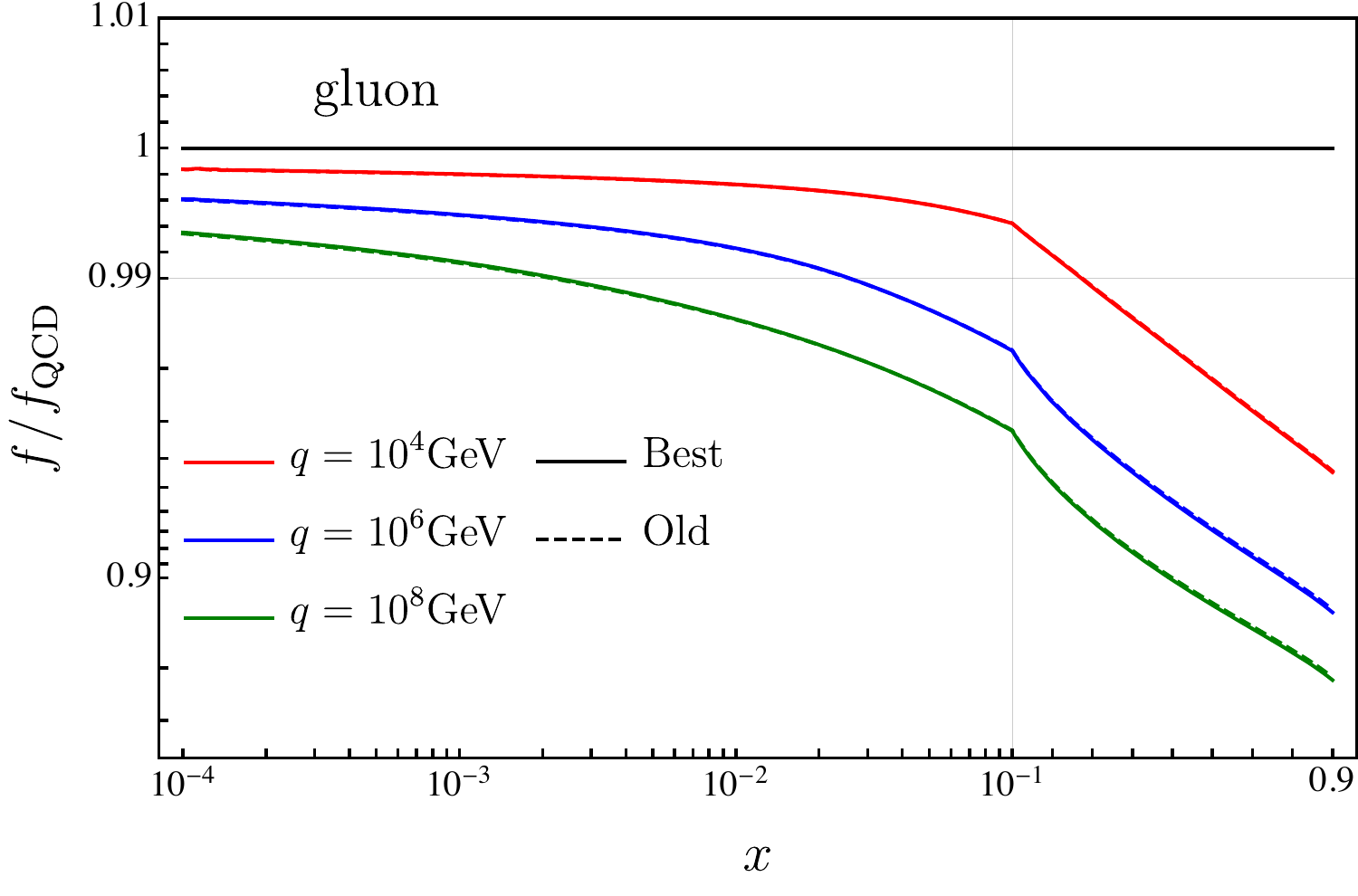}
   \caption{\label{fig:gluon}%
Gluon PDFs in the full unbroken SM, divided by their values assuming
pure QCD evolution only.  The thin gray lines show where the scales on the 
x- and/or y-axes switch between linear and logarithmic.}
}
We begin by showing resulting PDFs of strongly interacting particles. 
Figures~\ref{fig:gluon}, ~\ref{fig:quarksRight} and~\ref{fig:quarksLeft} show the evolution of the 
gluon, and well as 
left- and right-handed quark PDFs, 
normalized to their values assuming pure QCD evolution.
In each plot we show the results at three different scales, namely $q
= 10^4$, $10^6$ and $10^8$ GeV. The values of
$10^6$ and $10^8$ GeV are of course far away from energy scales one can
reach at any collider in the near or distant future. However, showing
the results at such unattainable values helps to illustrate their
approach to asymptotic behavior.

The improvements in this paper affect the gluon PDF at a level too small to be 
noticeable in Figure~\ref{fig:gluon}. This is expected because the gluon is overwhelmingly 
dominated by QCD evolution, and is only affected by electroweak corrections through 
the back-reaction from quarks. 

\FIGURE[h]{
 \centering
  \includegraphics[scale=0.45]{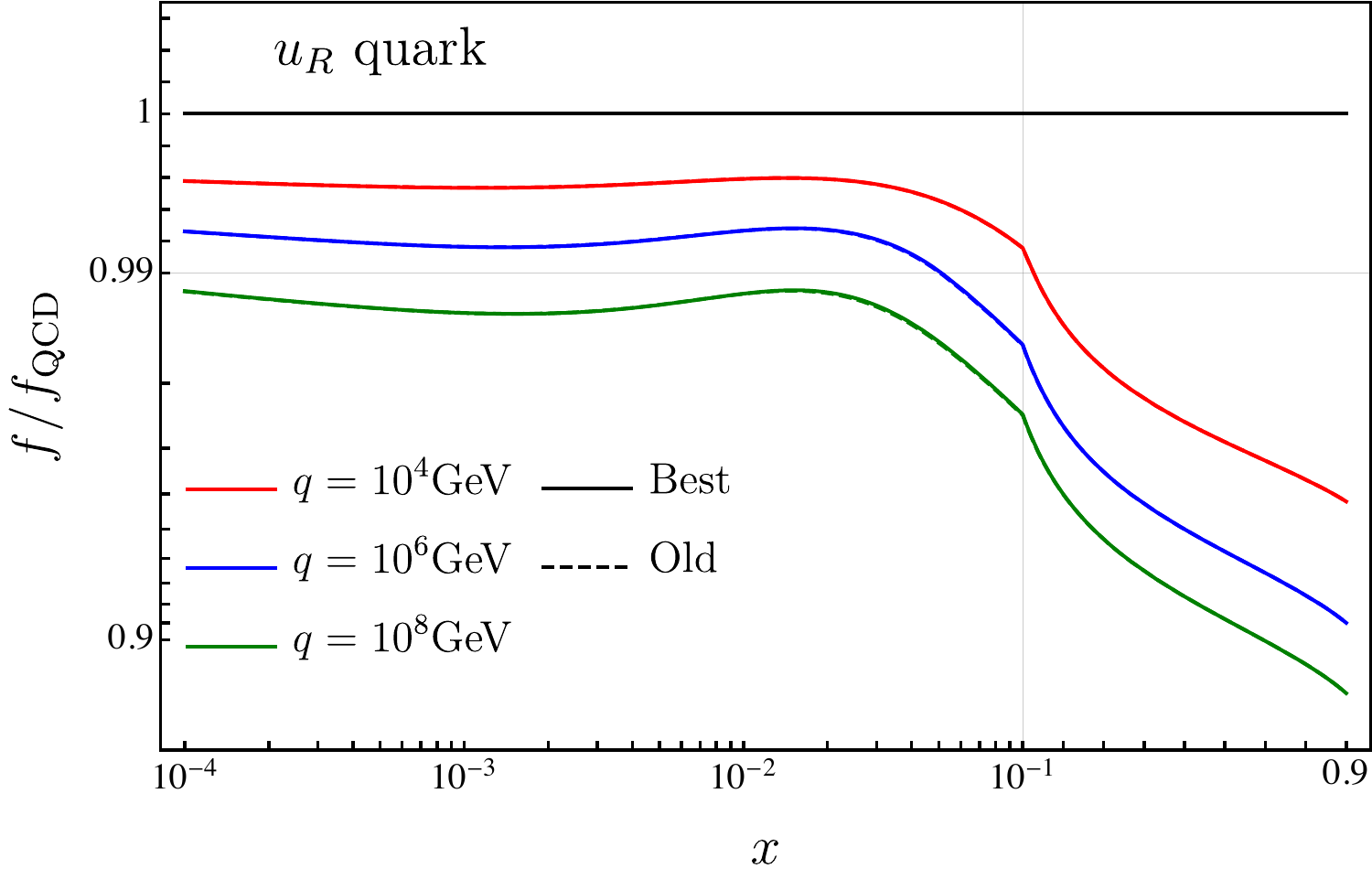}
  \includegraphics[scale=0.45]{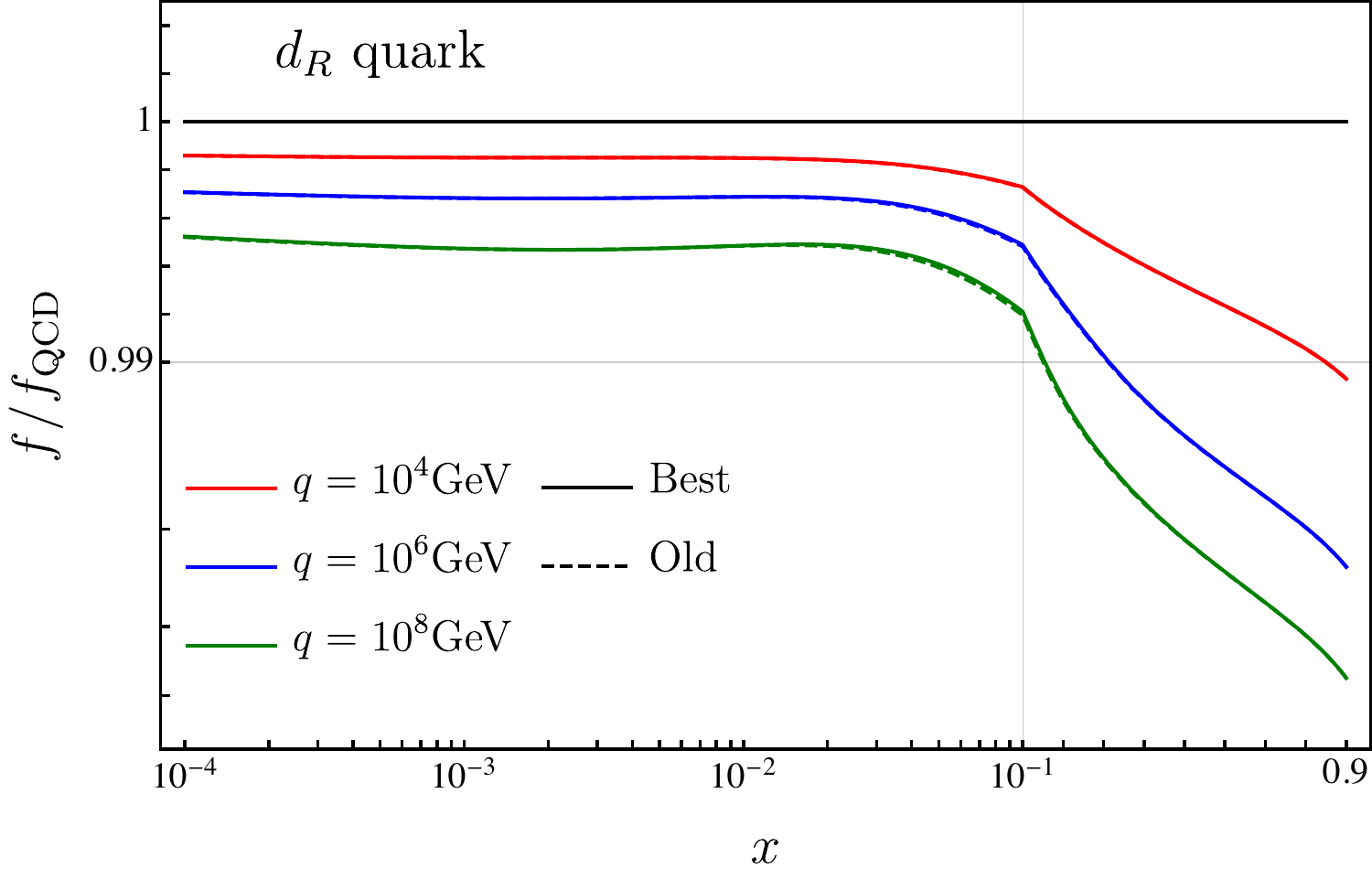}
  \includegraphics[scale=0.45]{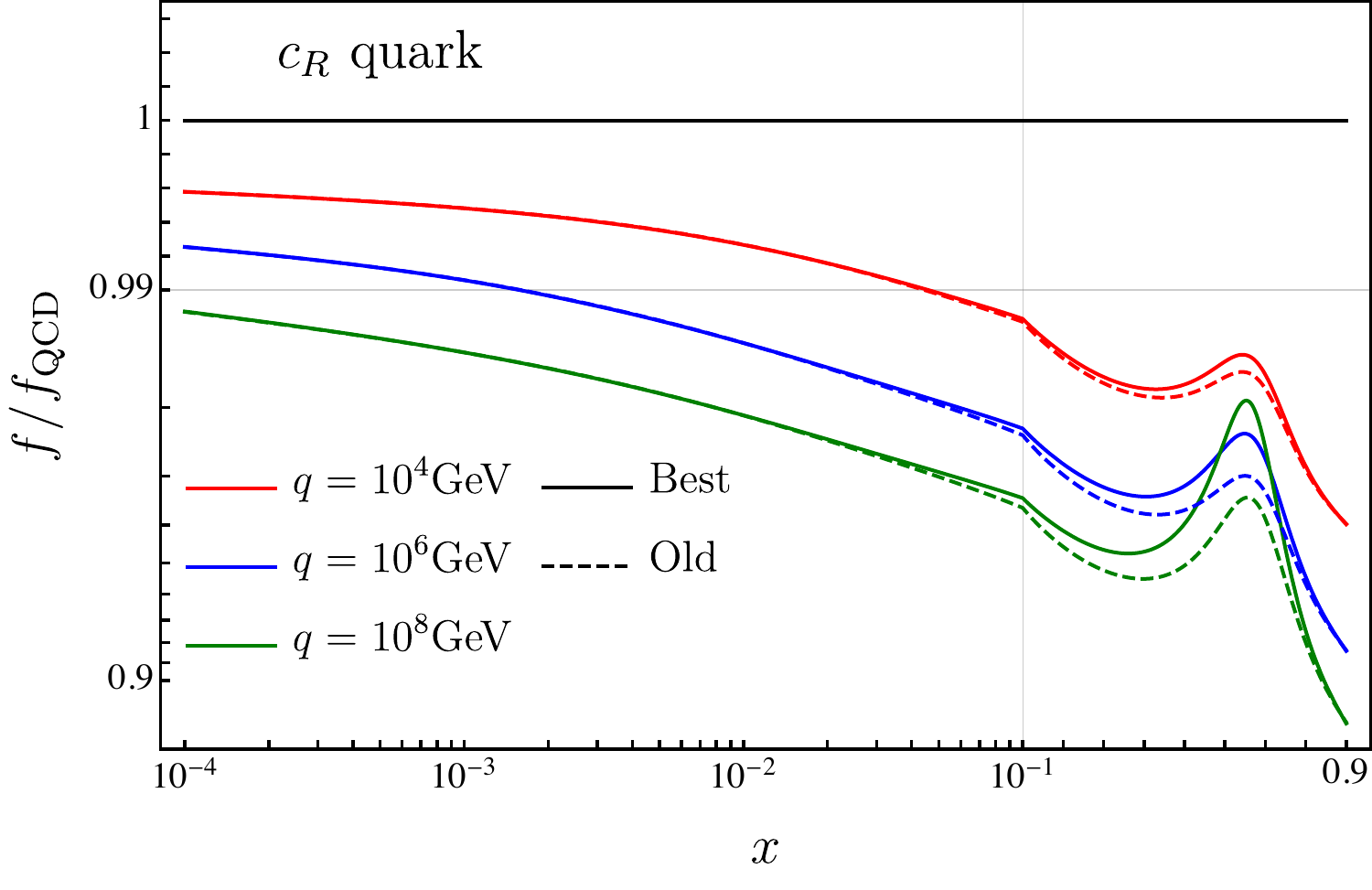}
  \includegraphics[scale=0.45]{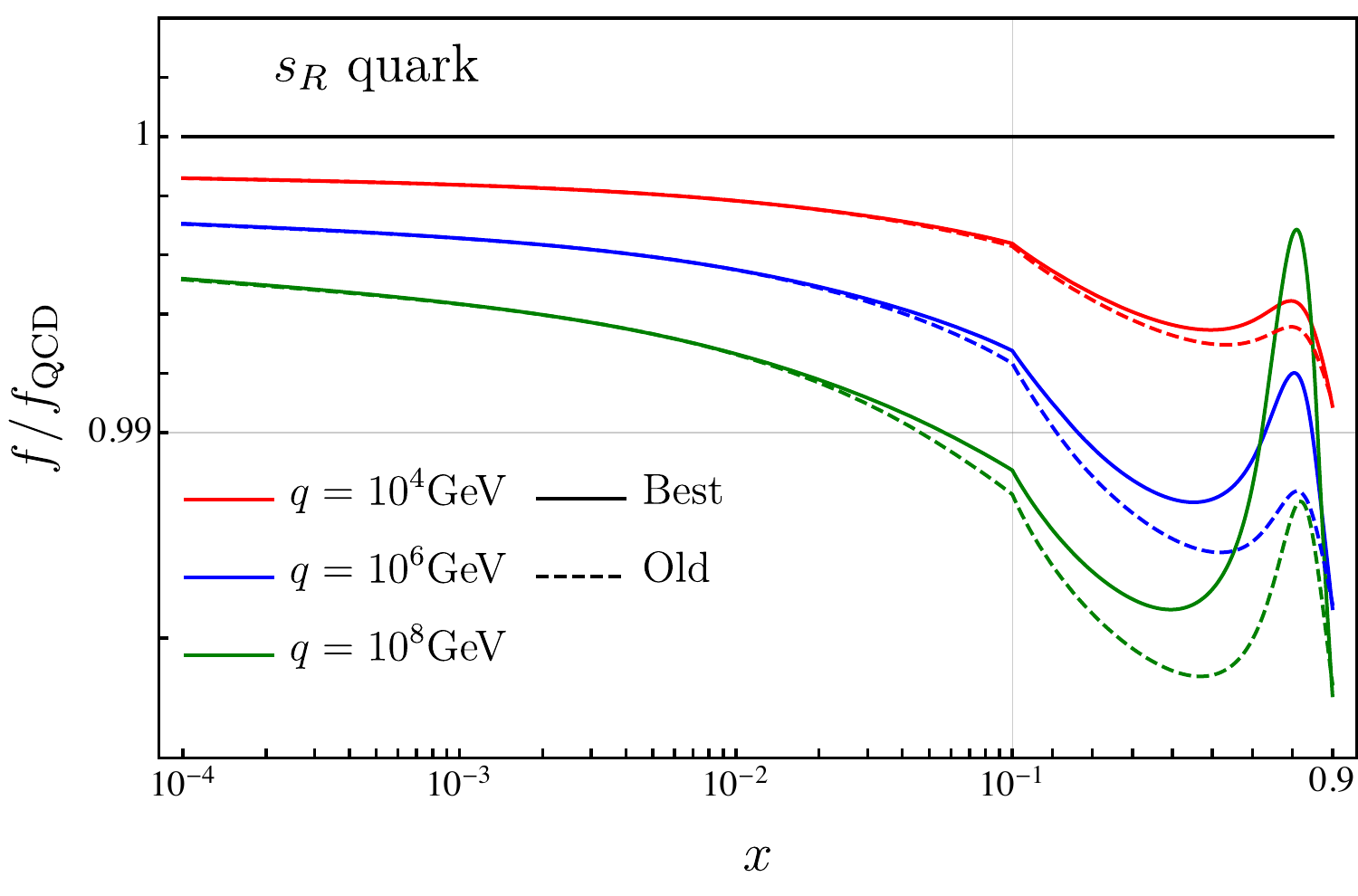}
  \includegraphics[scale=0.45]{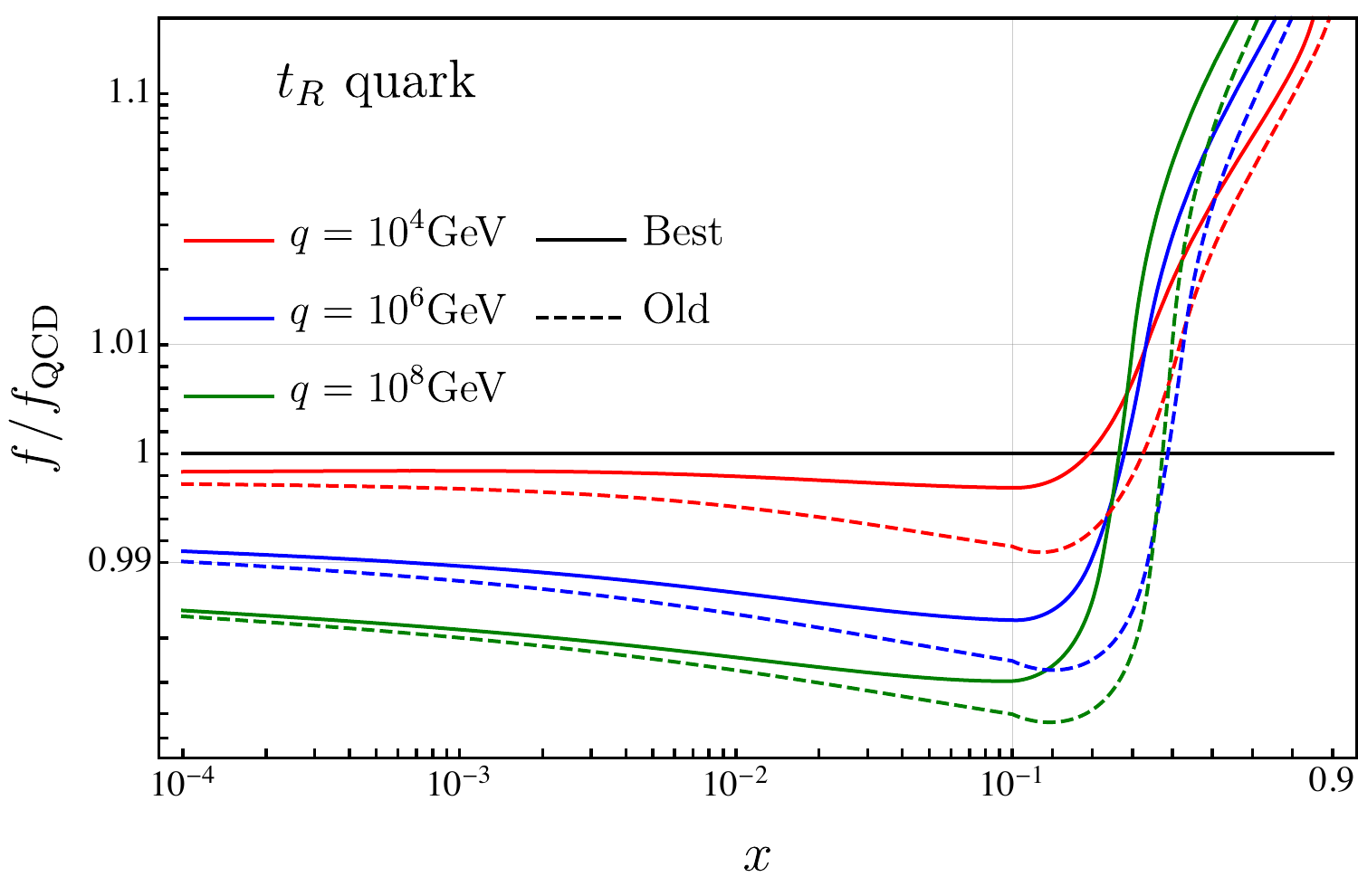}
  \includegraphics[scale=0.45]{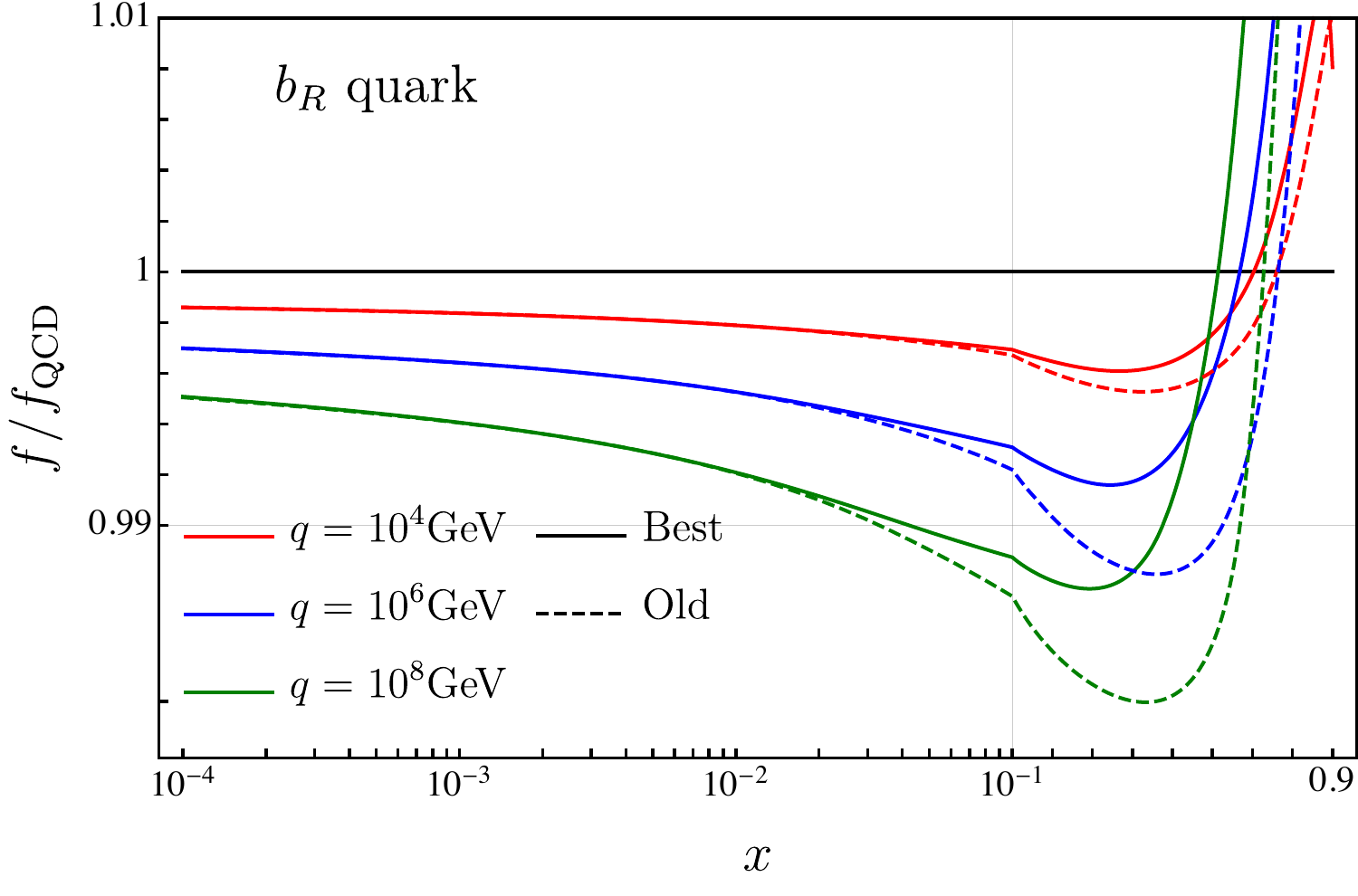}
   \caption{\label{fig:quarksRight}%
Right-handed quark PDFs in the full unbroken SM, divided by their values assuming
pure QCD evolution only.  The thin gray lines show where the scales on the 
x- and/or y-axes switch between linear and logarithmic.}
}
The right-handed quark PDFs have no double-logarithmic component and
mainly evolve to slightly lower values than pure QCD, due to energy loss through
the additional splitting $q_R \to q_R B$. The improvements of this paper affect the PDFs only 
at high $x$ and are much more pronounced for the heavy quarks. This is because heavy quarks 
are mainly produced perturbatively in QCD, such that the relative electroweak effect is overall larger. 

\FIGURE[h]{
 \centering
  \includegraphics[scale=0.45]{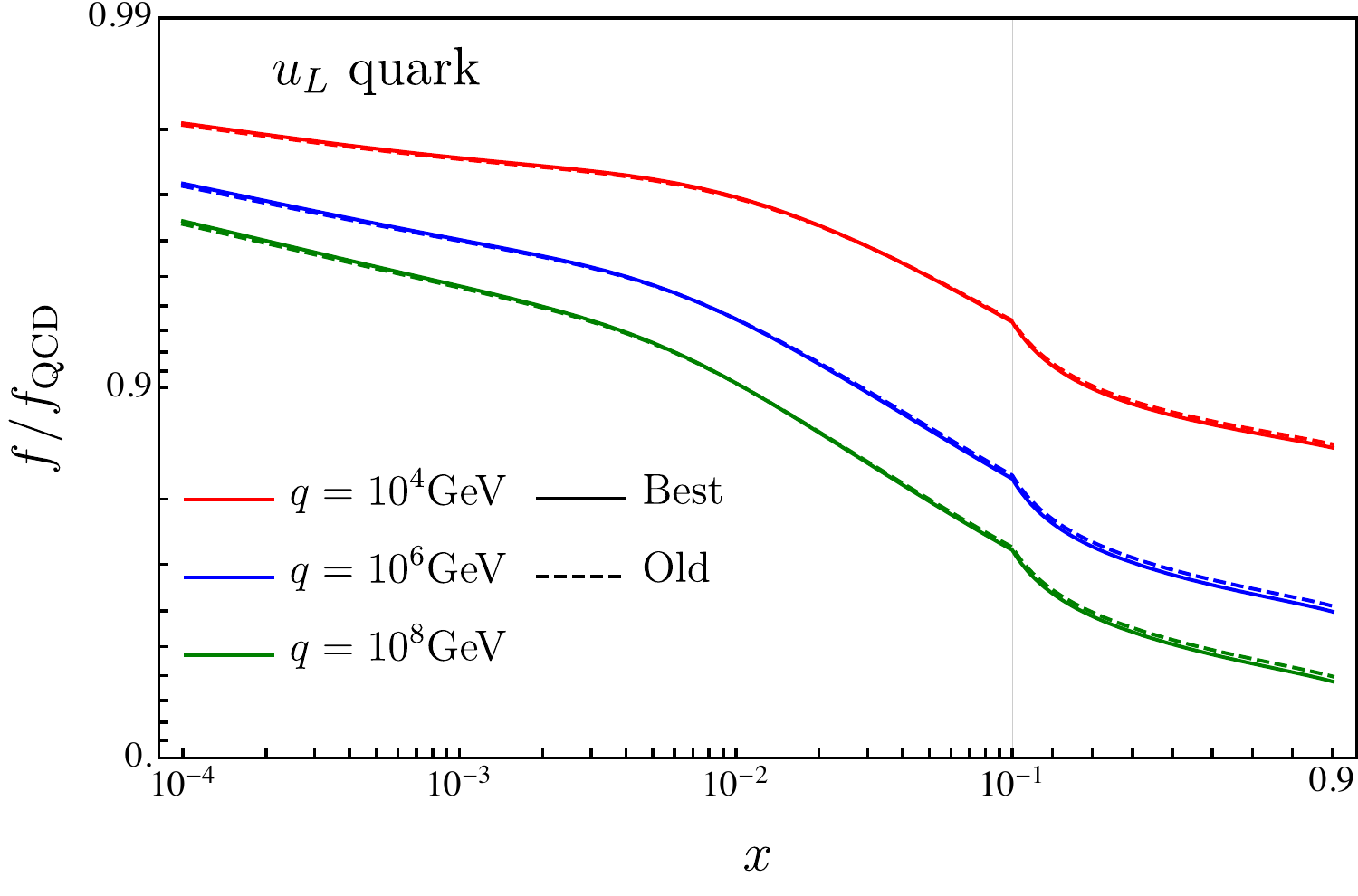}
  \includegraphics[scale=0.45]{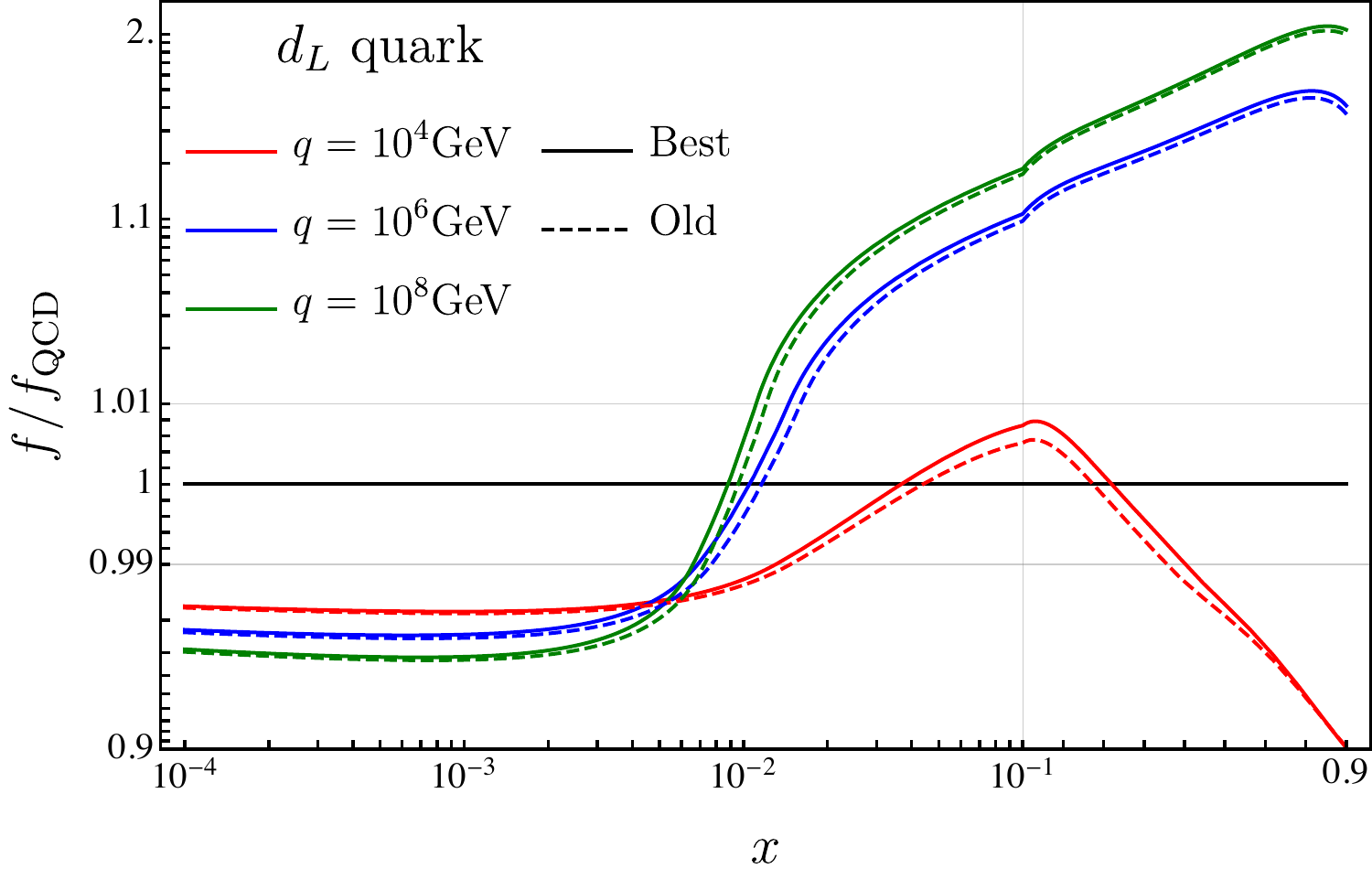}
  \includegraphics[scale=0.45]{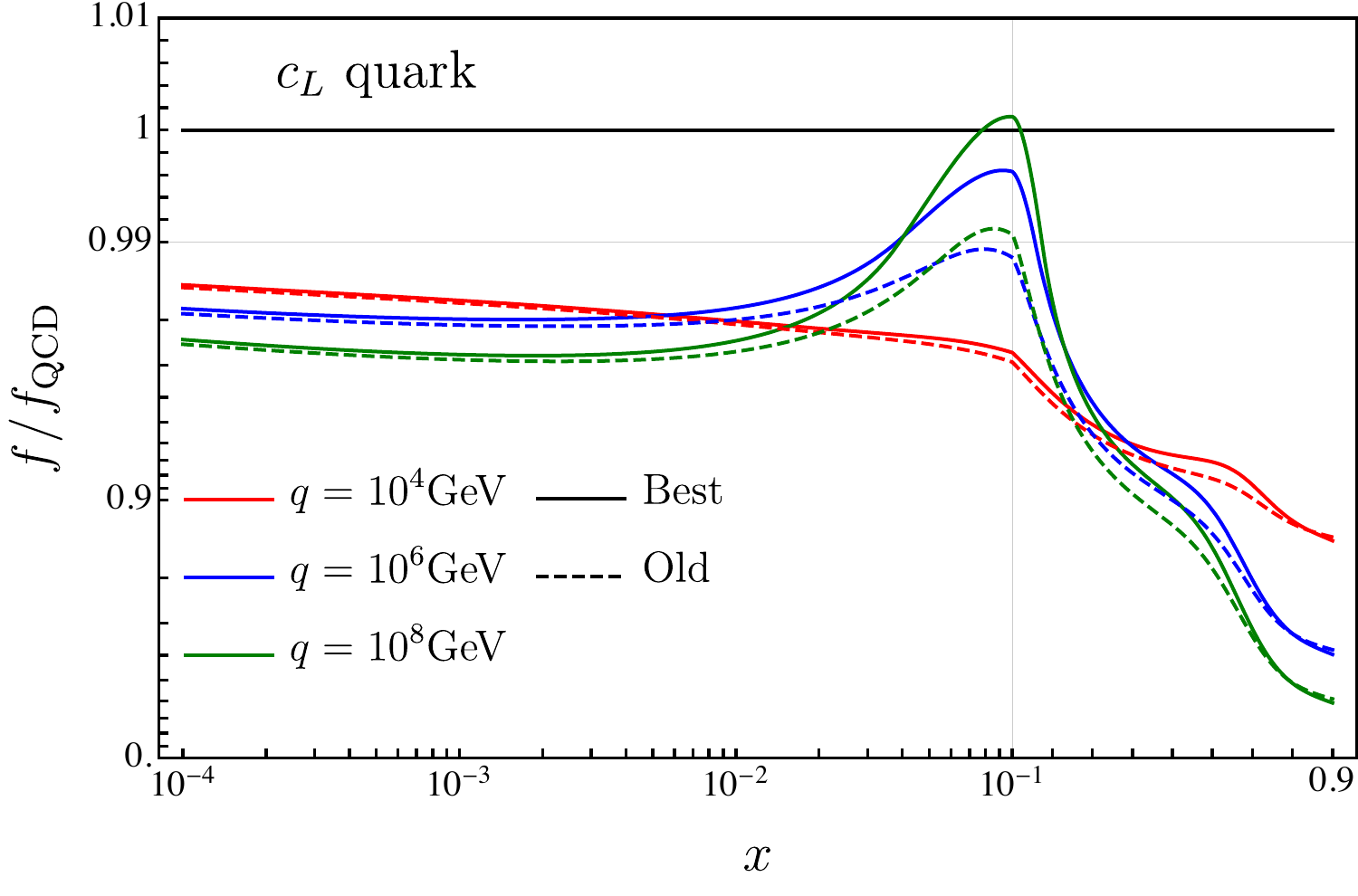}
  \includegraphics[scale=0.45]{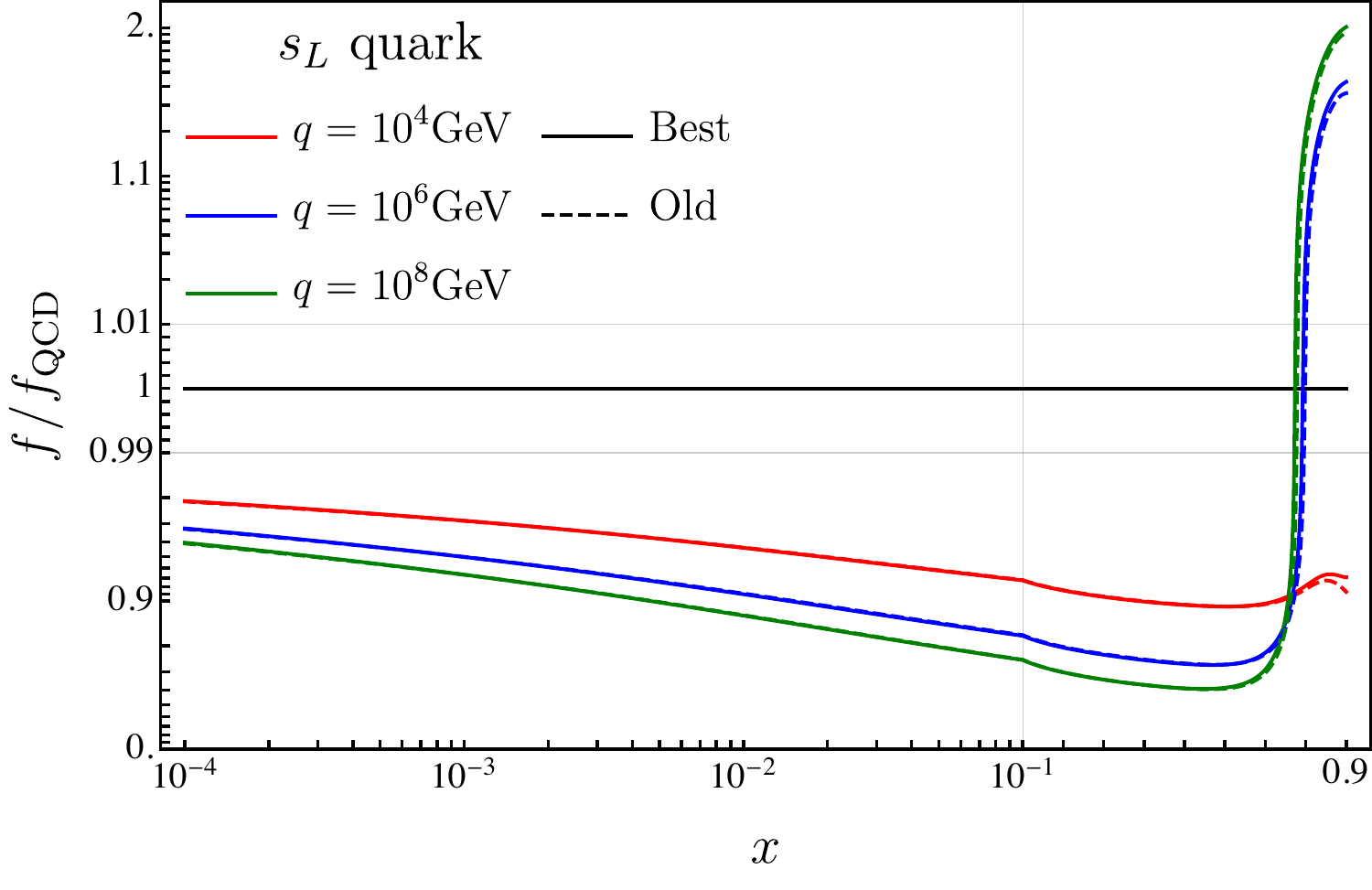}
  \includegraphics[scale=0.45]{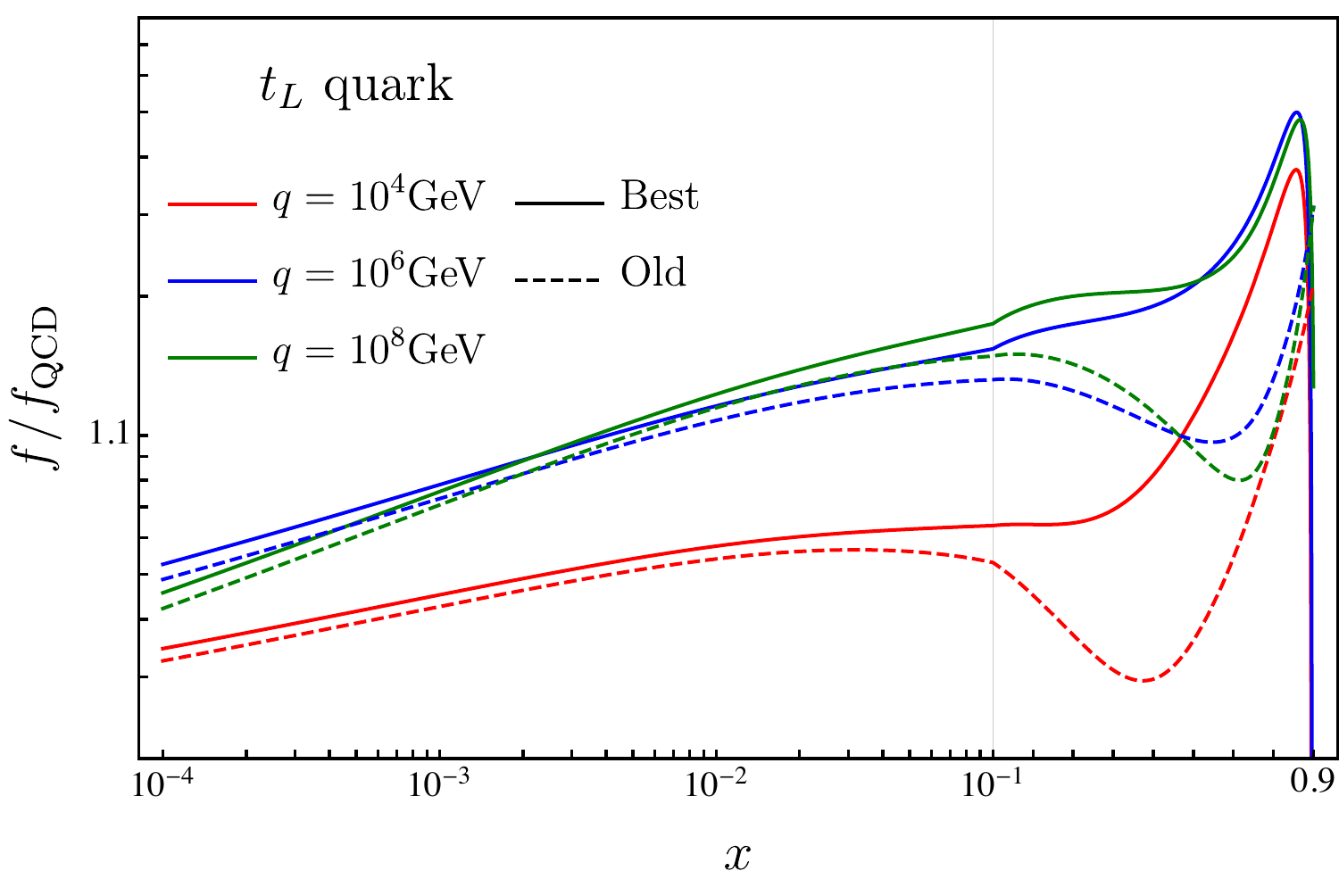}
  \includegraphics[scale=0.45]{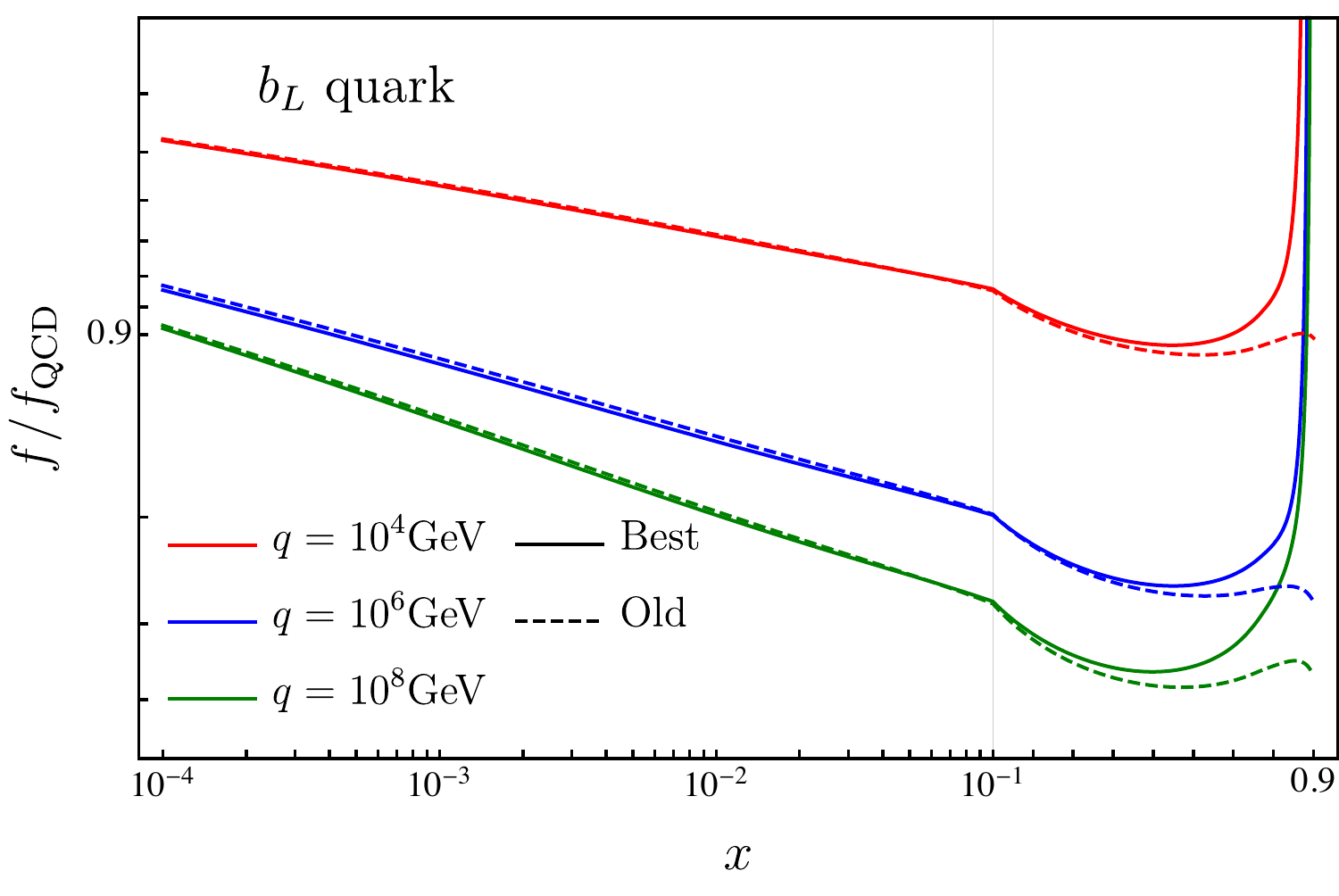}
   \caption{\label{fig:quarksLeft}%
Left-handed quark PDFs in the full unbroken SM, divided by their values assuming
pure QCD evolution only.  The thin gray lines show where the scales on the 
x- and/or y-axes switch between linear and logarithmic.}
}
For left-handed quarks, at low $x$, the effects of the improvements of
this paper are very small. As discussed in~\cite{Bauer:2017isx},  the
light quarks (and antiquarks, not shown) evolve to lower values
compared to pure QCD at small $x$, due to an overall loss of energy to
the electroweak gauge bosons.
At large $x$, the effects are more noticeable, and 
in particular for the heavy quarks lead to ${\cal O}(1)$ relative changes,
although the absolute values of the PDFs there are very small.
The qualitative features are unchanged, in particular the up and down
quarks (top row) exhibit different behaviors, with the left-handed up
PDF evolving more rapidly to lower values compared to pure QCD, while
the down quark eventually evolves to higher values, as the isovector
contribution to their PDFs dies away double-logarithmically.

\FIGURE{
	\centering
	\includegraphics[scale=0.45]{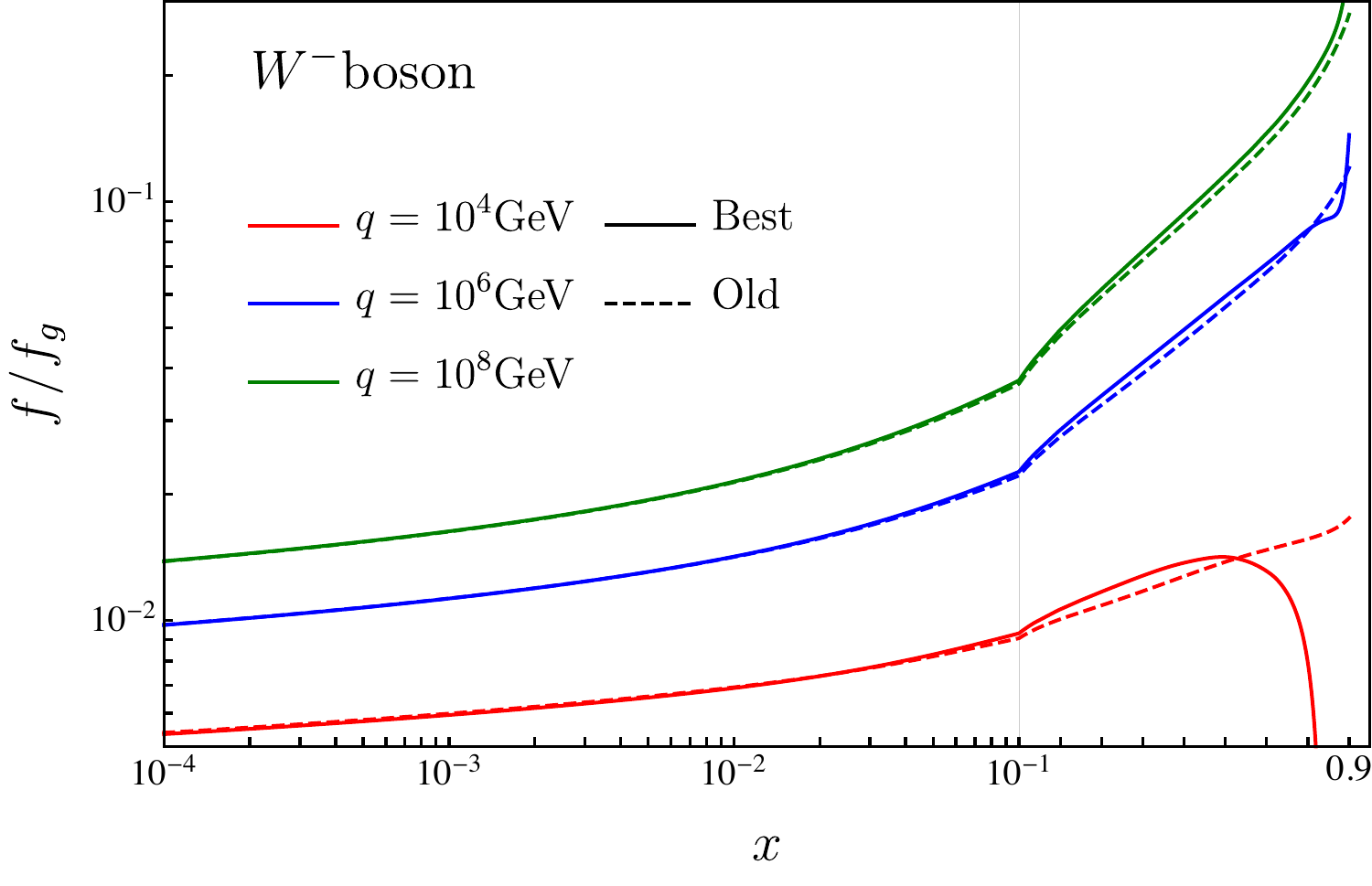}
	\includegraphics[scale=0.45]{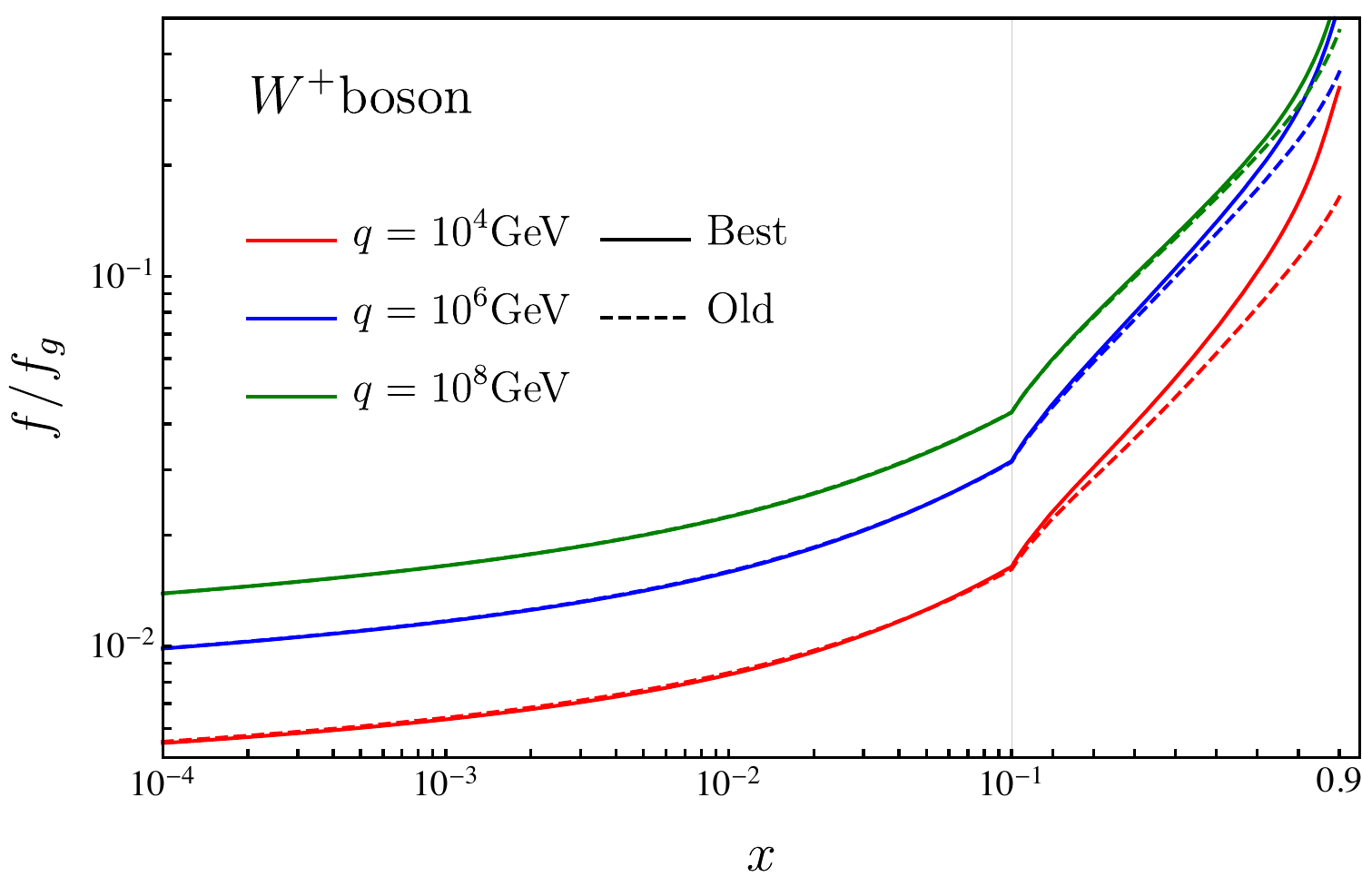}
	\includegraphics[scale=0.45]{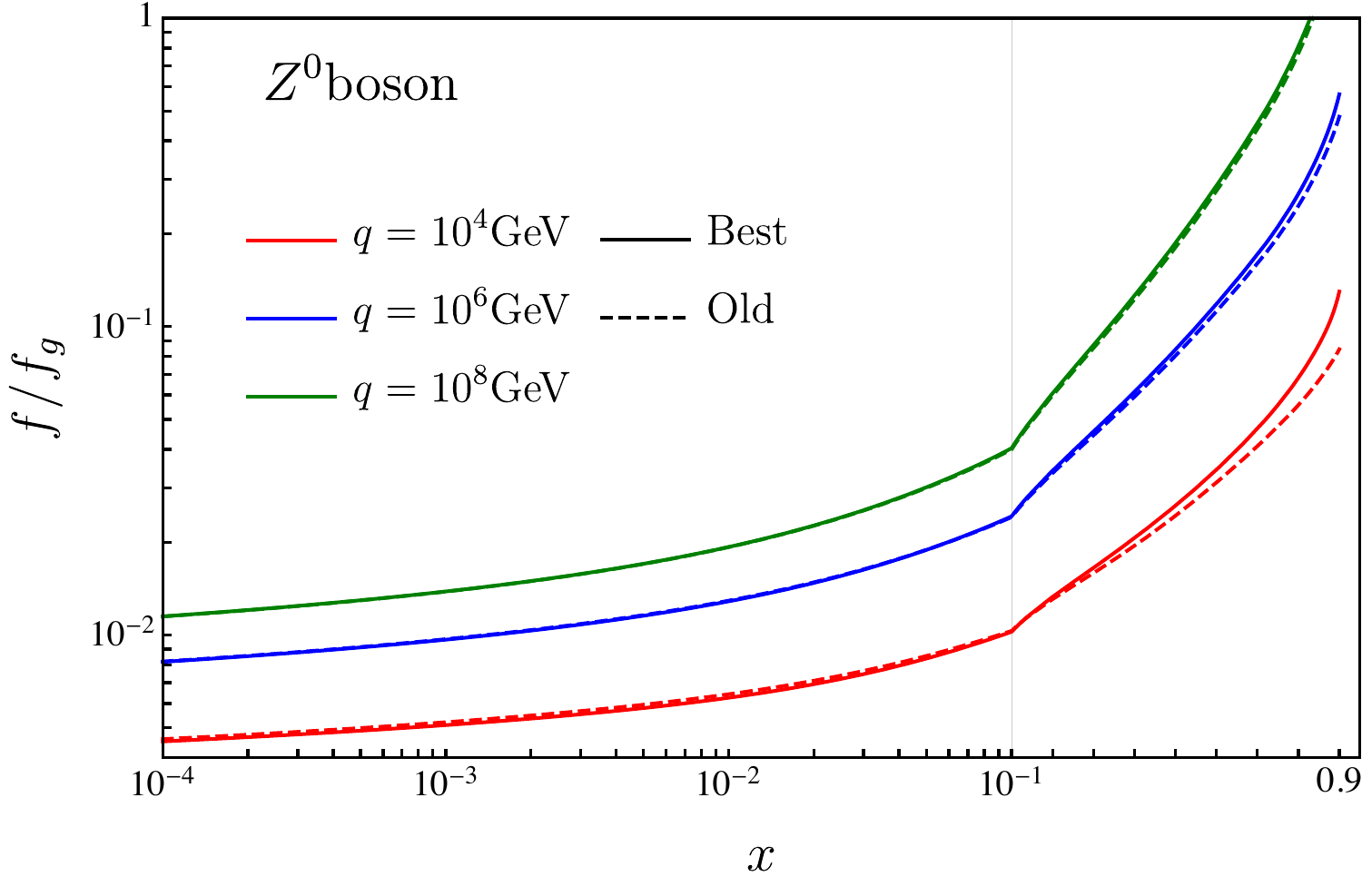}
	\includegraphics[scale=0.45]{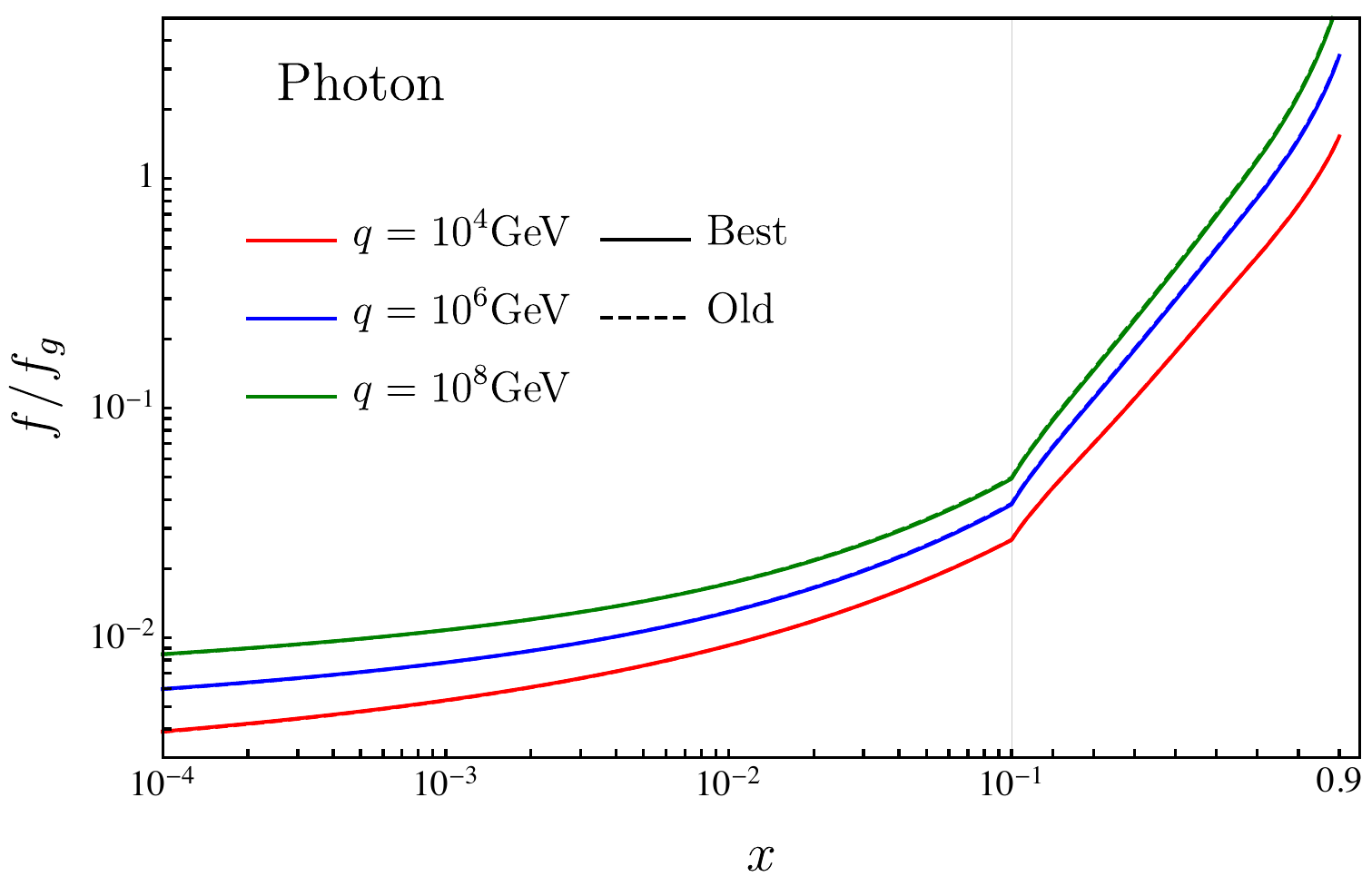}
	\includegraphics[scale=0.45]{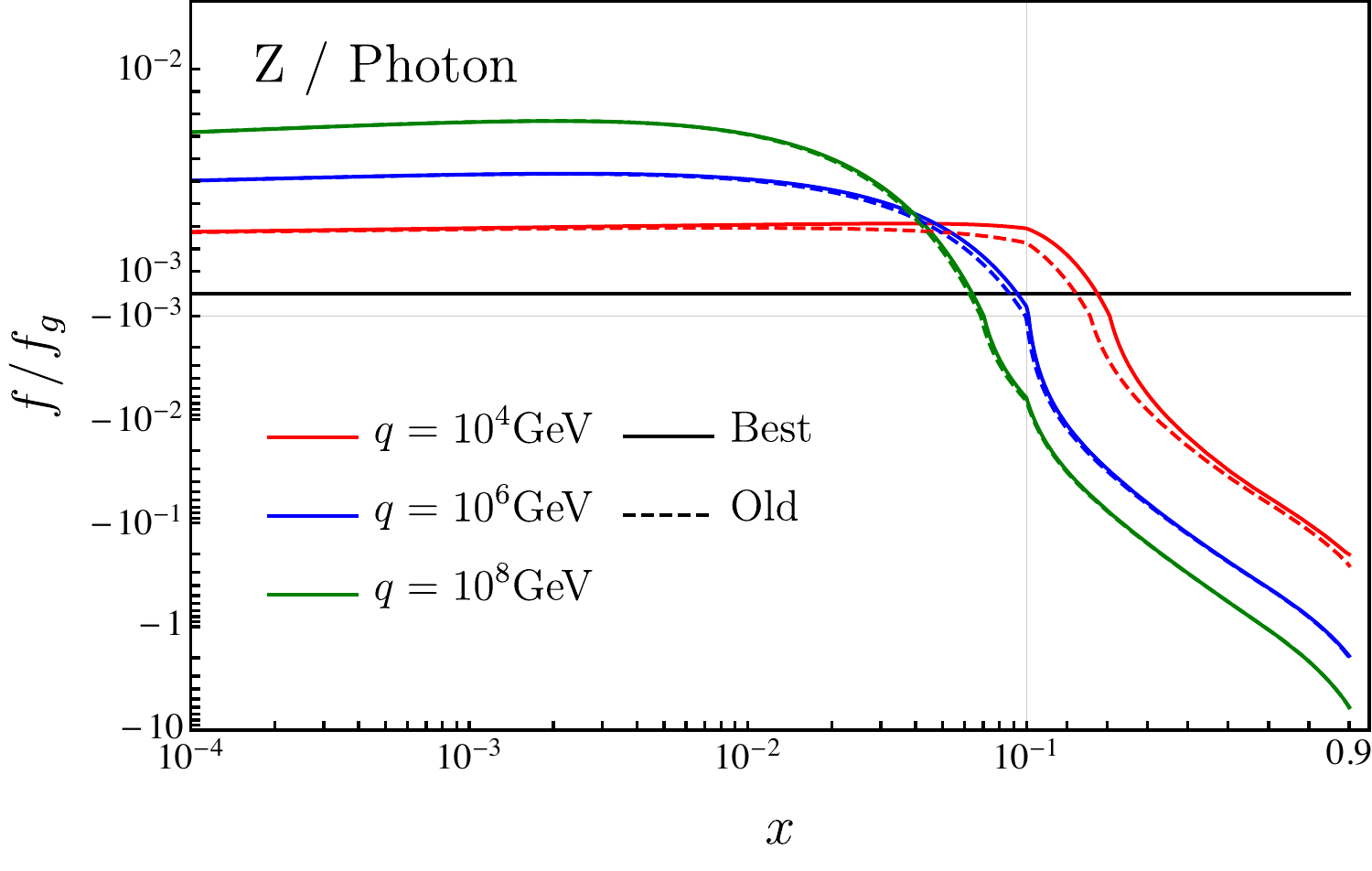}
	\caption{\label{fig:VectorBosons}%
		Unpolarized transverse electroweak boson PDFs normalized by the gluon PDF.  The thin gray lines show where the scales on the 
x- and/or y-axes switch between linear and logarithmic.
	}
}
\FIGURE{
	\centering
	\includegraphics[scale=0.45]{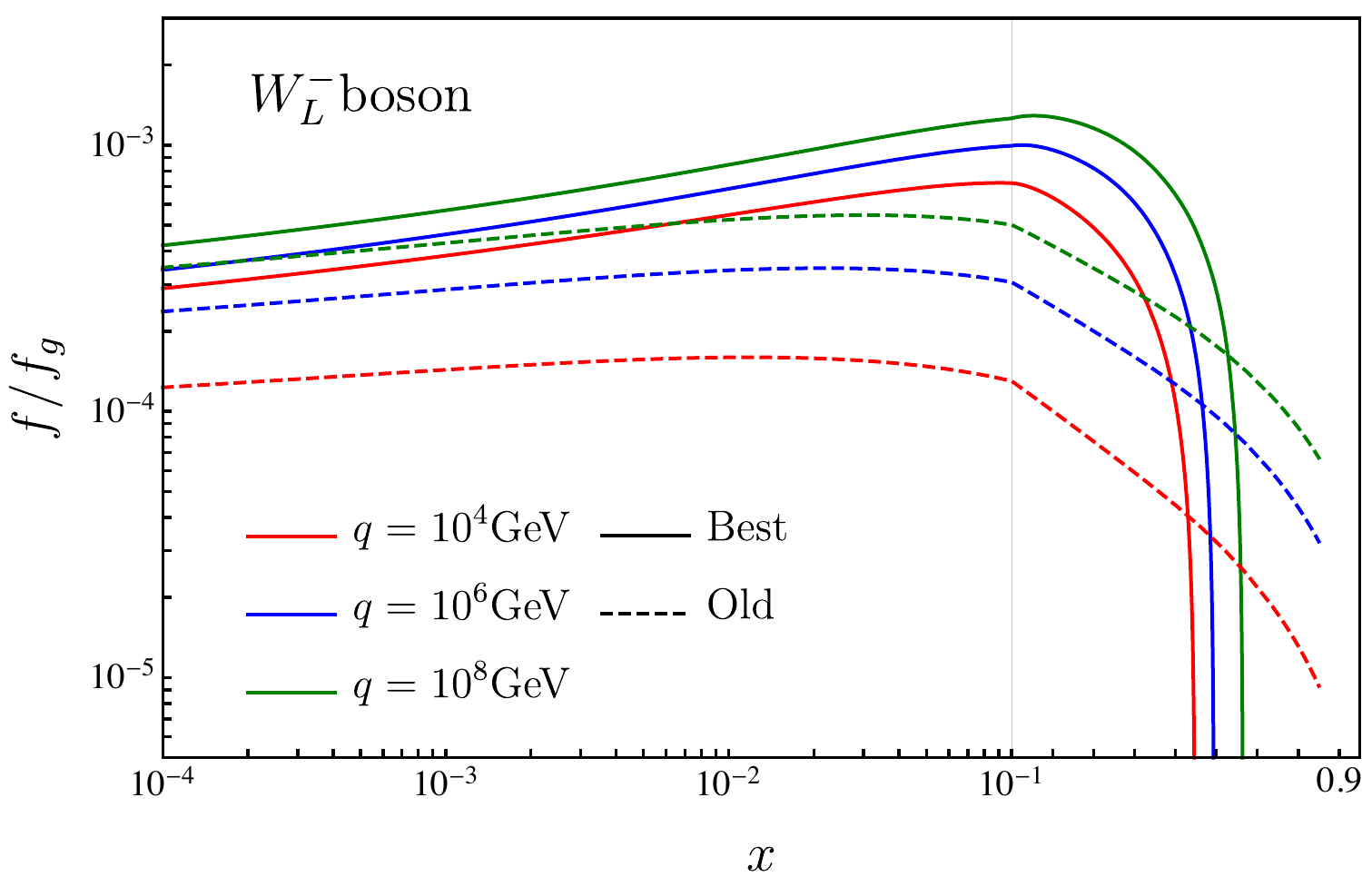}
	\includegraphics[scale=0.45]{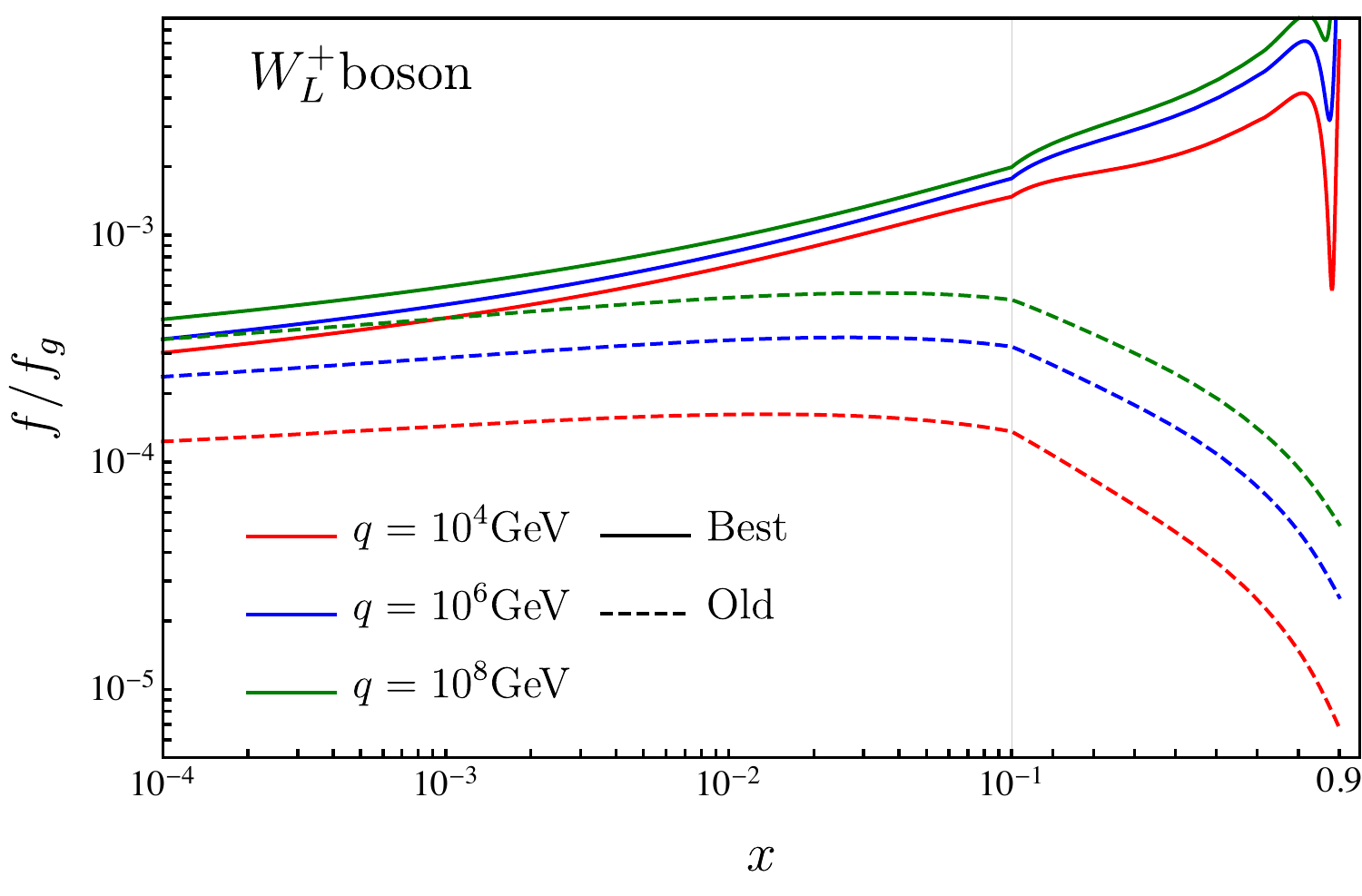}
	\includegraphics[scale=0.45]{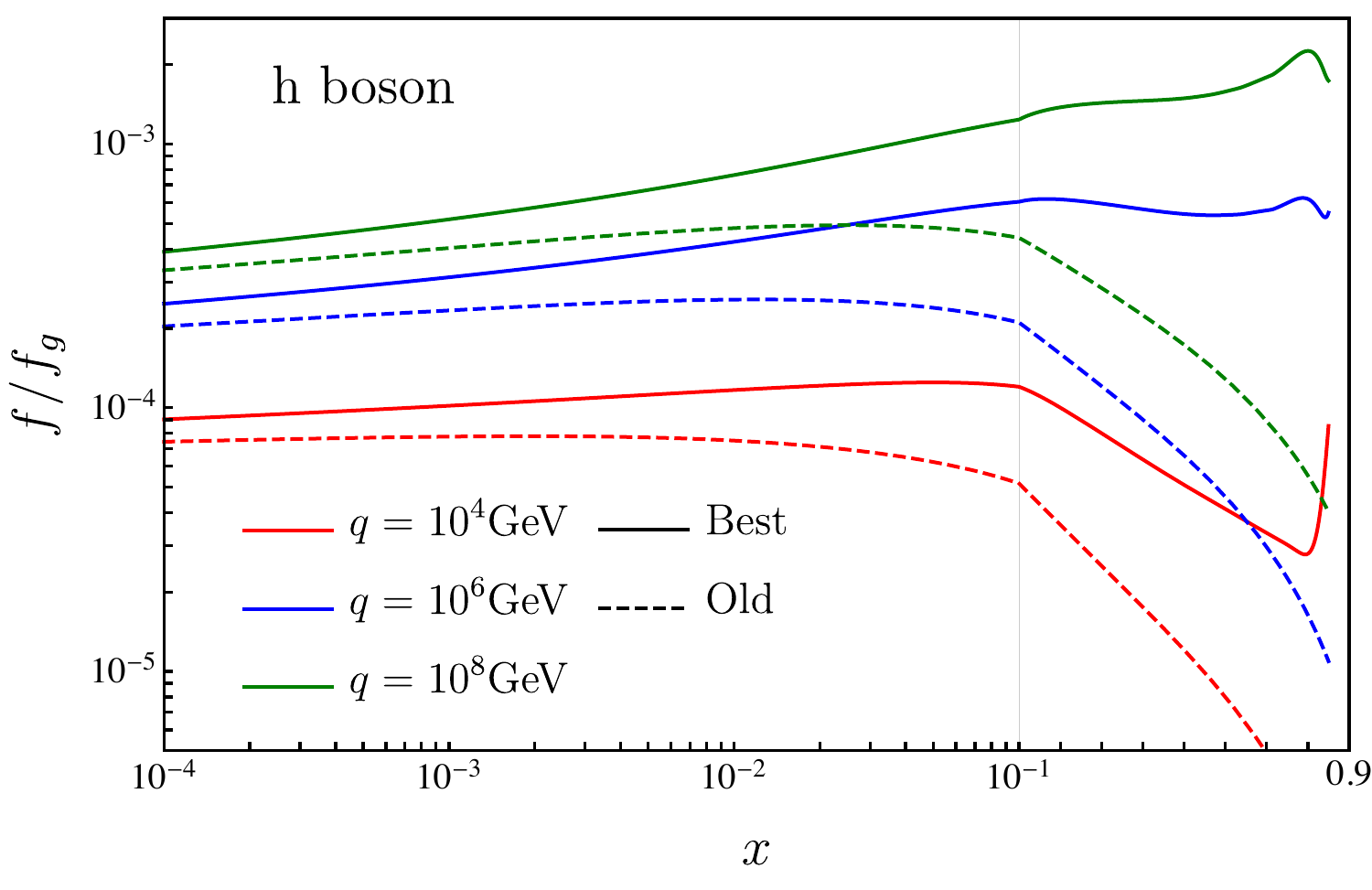}
	\includegraphics[scale=0.45]{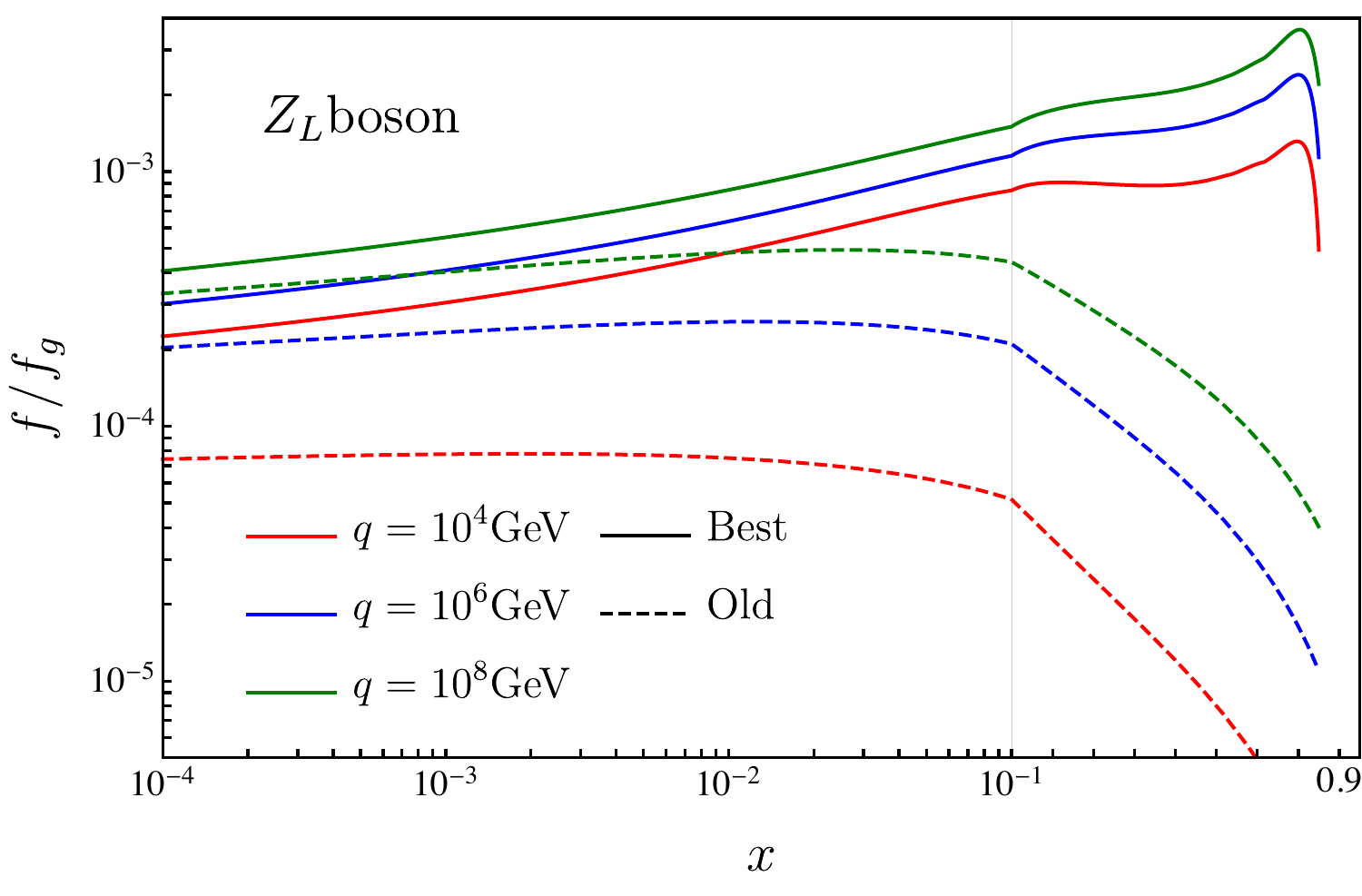}
	\includegraphics[scale=0.45]{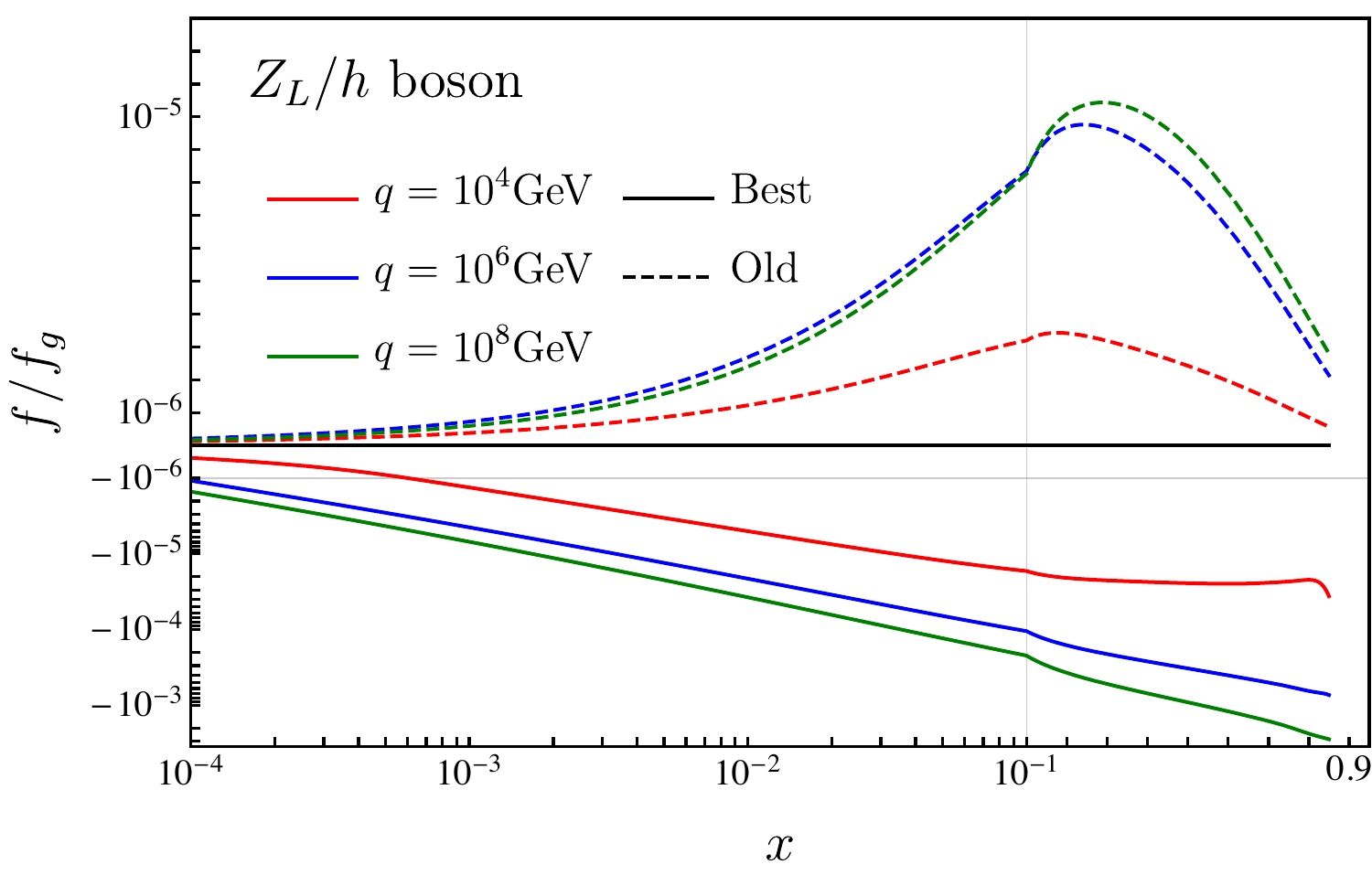}
	\caption{\label{fig:HiggsBosons}%
		 Longitudinal gauge and Higgs boson PDFs normalized by the gluon PDF. The $Z_L/h$ PDF is purely imaginary and we show the result divided by $i$. The thin gray line shows where the scales on the 
x- and/or y-axes switch between linear and logarithmic.
	}
}
Next, we study the effect on vector boson PDFs. Recall that in~\cite{Bauer:2017isx,Bauer:2017bnh} the initial values 
for the heavy gauge bosons at the matching scale $q_0$ were zero and
their entire effect was generated dynamically through the DGLAP
evolution above that scale. In contrast,  in this work we use the
results of~\cite{Fornal:2018znf} to determine their initial
values. These input values are ${\cal O}(\alpha)$  and thus of
subleading logarithmic order. At relatively low values of $q$ we
therefore expect large effects, while at large $q$ values
the logarithmic corrections should dominate, such that the effect of
the input decreases. This can be seen clearly in Fig.~\ref{fig:VectorBosons},
where we show the ratio of the PDFs relative to the gluon. 
Since we did not change the initial condition of the photon, its PDF is not affected. For the heavy vector boson PDFs the effect is 
more pronounced at low values of $q$ and is barely noticeable at the largest value of $q$ shown. 

For the longitudinally polarized gauge bosons,
the Higgs boson and the mixed PDF between the Higgs and the $Z_L$, the effect of the improvements is considerable larger, and at large
$x$ changes the PDFs by more than an order of magnitude. This is 
because their contributions from the dynamical evolution are much smaller,  arising only to second order in the electroweak gauge coupling, 
and through Yukawa couplings to the top quark. The initial values, on the other hand are of the same order as for the transverse 
vector bosons, namely ${\cal O}(\alpha)$. This can be traced back to the fact that the equivalence theorem, which underlies the 
DGLAP evolution in the unbroken SM, is badly broken at scales of order
of the electroweak scale, manifesting itself through power corrections
that are large at threshold (see also \cite{Chen:2016wkt}). By using the perturbative result as the
initial value to the DGLAP evolution, one combines these large threshold
corrections with the large logarithmic terms that dominate far above
the threshold.

To illustrate the uncertainties associated with subleading terms, we
show in Tables~\ref{tab:momfracs10} and \ref{tab:momfracs100} the
dependence of some integrated PDFs (momentum fractions) on the
infrared cutoff $m_V$ and matching scale $q_0$.  The electroweak PDFs
are much less sensitive to these parameters than was the case in
Ref.~\cite{Bauer:2017isx}, due to the electroweak input at the
matching scale.  The exception is the Higgs boson, which is still
generated dynamically starting from zero at the matching scale.
\begin{table}[h!]
\begin{center}
\begin{tabular}{|c|c|c|c|c|c|c|c|c|c|}
\hline
$m_V$/GeV & $q_0$/GeV & $u_L$ & $t_L$ & $W_T^+$ & $W_T^-$ & $e_L^-$ & $\nu_e$
  & $h$ & $Z_L^0$ \\\hline
100 & 100 & 8.51 & 0.43 & 0.46 & 0.34 & 0.0021 & 0.0014 & 0.0044 & 0.0232\\
  50 & 100 & 8.42 & 0.44 & 0.46 & 0.34 & 0.0020 & 0.0014 & 0.0053 & 0.0233\\
  50 & 200 & 8.48 & 0.44 & 0.45 & 0.33 & 0.0020 & 0.0013 & 0.0051 & 0.0230\\
100 & 200 & 8.57 & 0.43 & 0.45 & 0.32 & 0.0020 & 0.0013 & 0.0043 & 0.0230\\
200 & 200 & 8.64 & 0.42 & 0.45 & 0.32 & 0.0020 & 0.0013 & 0.0037 & 0.0231\\
\hline
\end{tabular}
\end{center}
\caption{\label{tab:momfracs10} Momentum fractions (\%) carried by
  various parton species at scale $q=10$ TeV.}
\end{table}

\begin{table}[h!]
\begin{center}
\begin{tabular}{|c|c|c|c|c|c|c|c|c|c|}
\hline
$m_V$/GeV & $q_0$/GeV & $u_L$ & $t_L$ & $W_T^+$ & $W_T^-$ & $e_L^-$ & $\nu_e$
  & $h$ & $Z_L^0$ \\\hline
100 & 100 & 7.52 & 0.60 & 0.64 & 0.50 & 0.0034 & 0.0029 & 0.0107 & 0.0251\\
  50 & 100 & 7.41 & 0.62 & 0.64 & 0.51 & 0.0034 & 0.0029 & 0.0118 & 0.0251\\
  50 & 200 & 7.46 & 0.62 & 0.63 & 0.50 & 0.0033 & 0.0028 & 0.0116 & 0.0249\\
100 & 200 & 7.57 & 0.60 & 0.63 & 0.50 & 0.0033 & 0.0028 & 0.0105 & 0.0250\\
200 & 200 & 7.67 & 0.59 & 0.63 & 0.49 & 0.0034 & 0.0027 & 0.0095 & 0.0250\\
\hline
\end{tabular}
\end{center}
\caption{\label{tab:momfracs100} Momentum fractions (\%) carried by
  various parton species at scale $q=100$ TeV.}
\end{table}

Finally, we show the size of the vector boson polarization generated
by the electroweak evolution in Fig.~\ref{fig:Polarization}. As already mentioned, polarized vector
bosons were not included in our previous results.  We can see that for
the massive electroweak gauge bosons the polarization is ${\cal
  O}(1)$, especially at large $x$, and negative owing to the dominance of
emission from left-handed fermions.
For the photon, and even more so the gluon, the polarization is much smaller.

\FIGURE{
	\centering
	\includegraphics[scale=0.45]{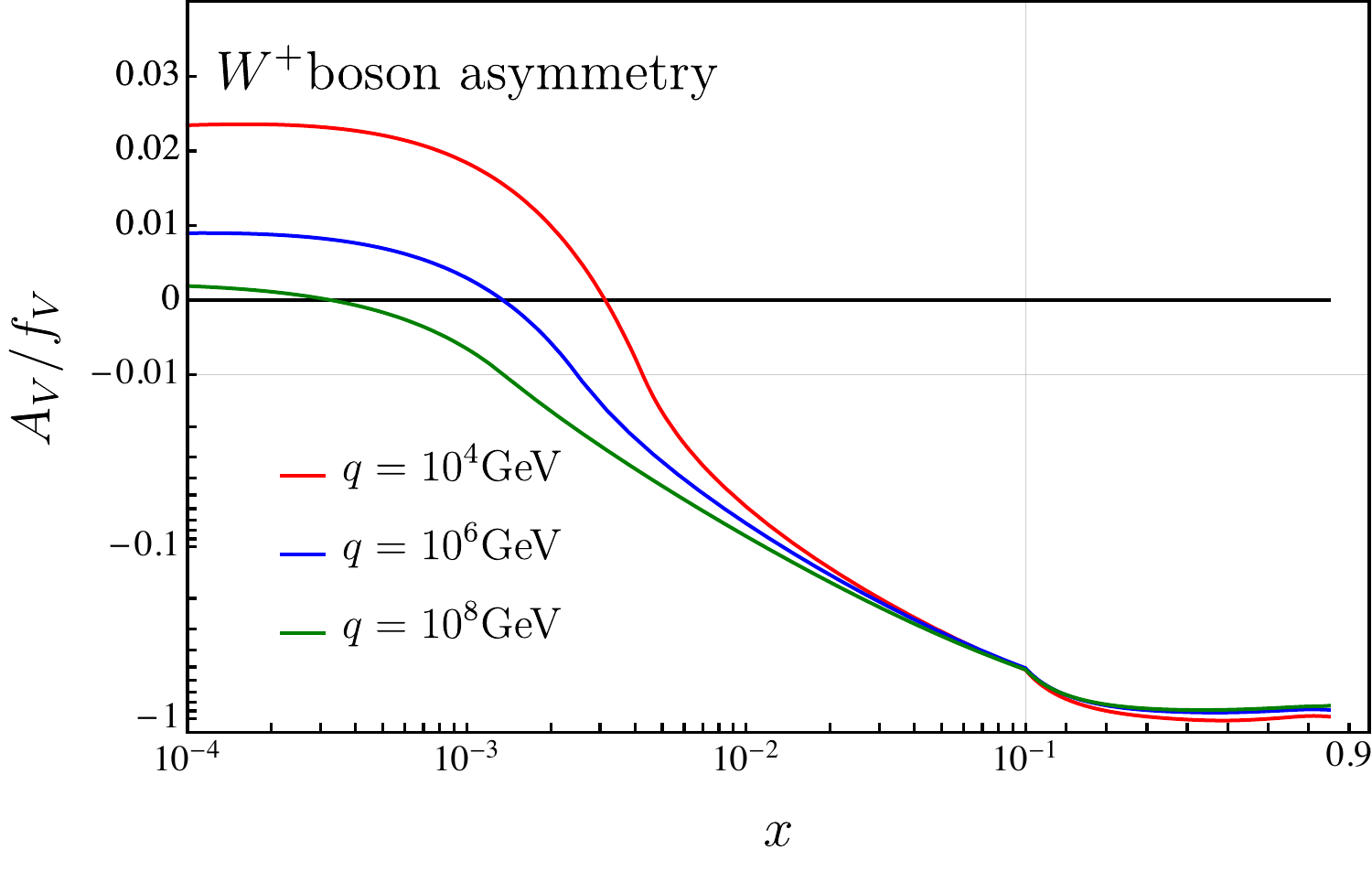}
	\includegraphics[scale=0.45]{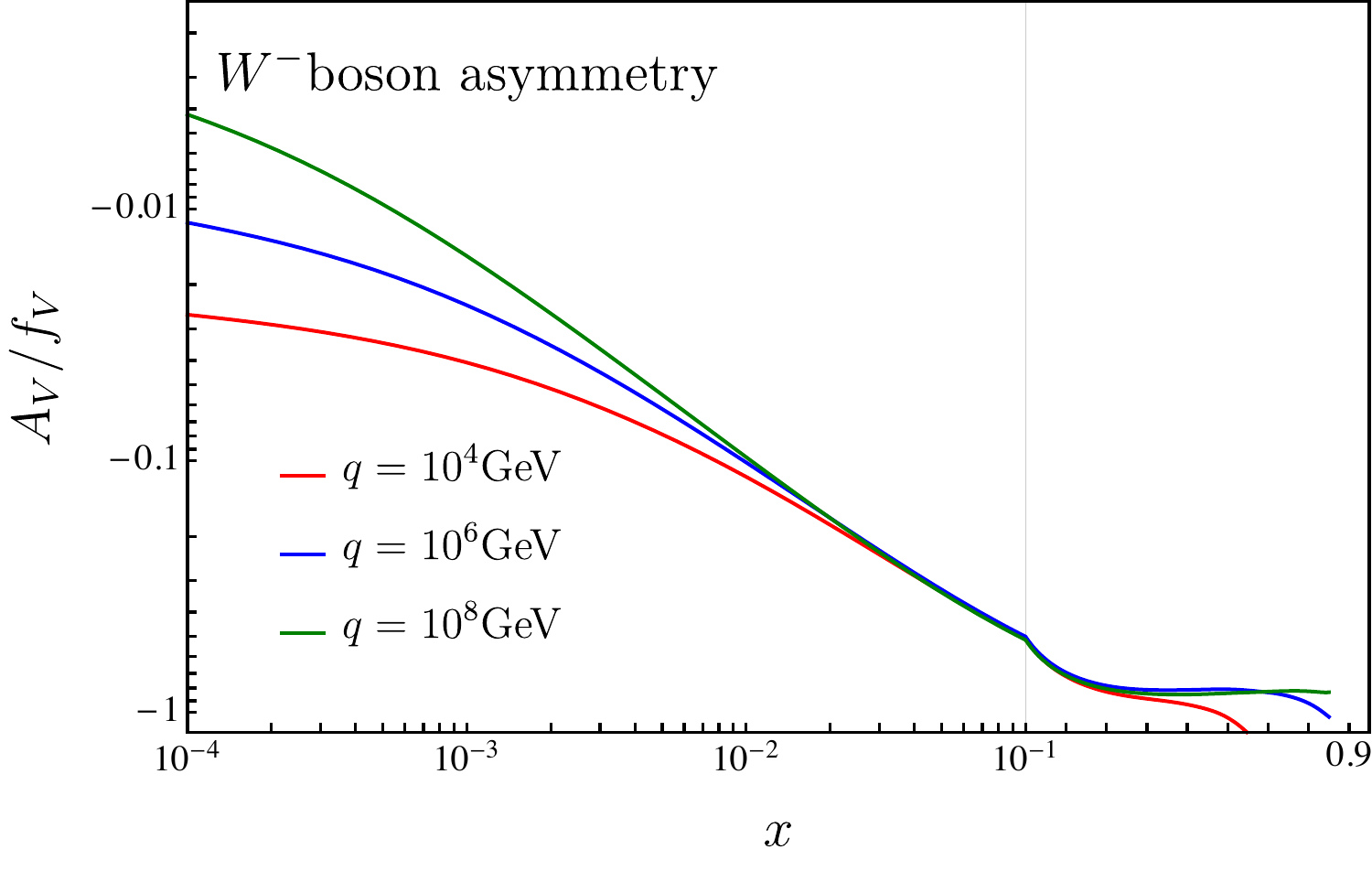}
	\includegraphics[scale=0.45]{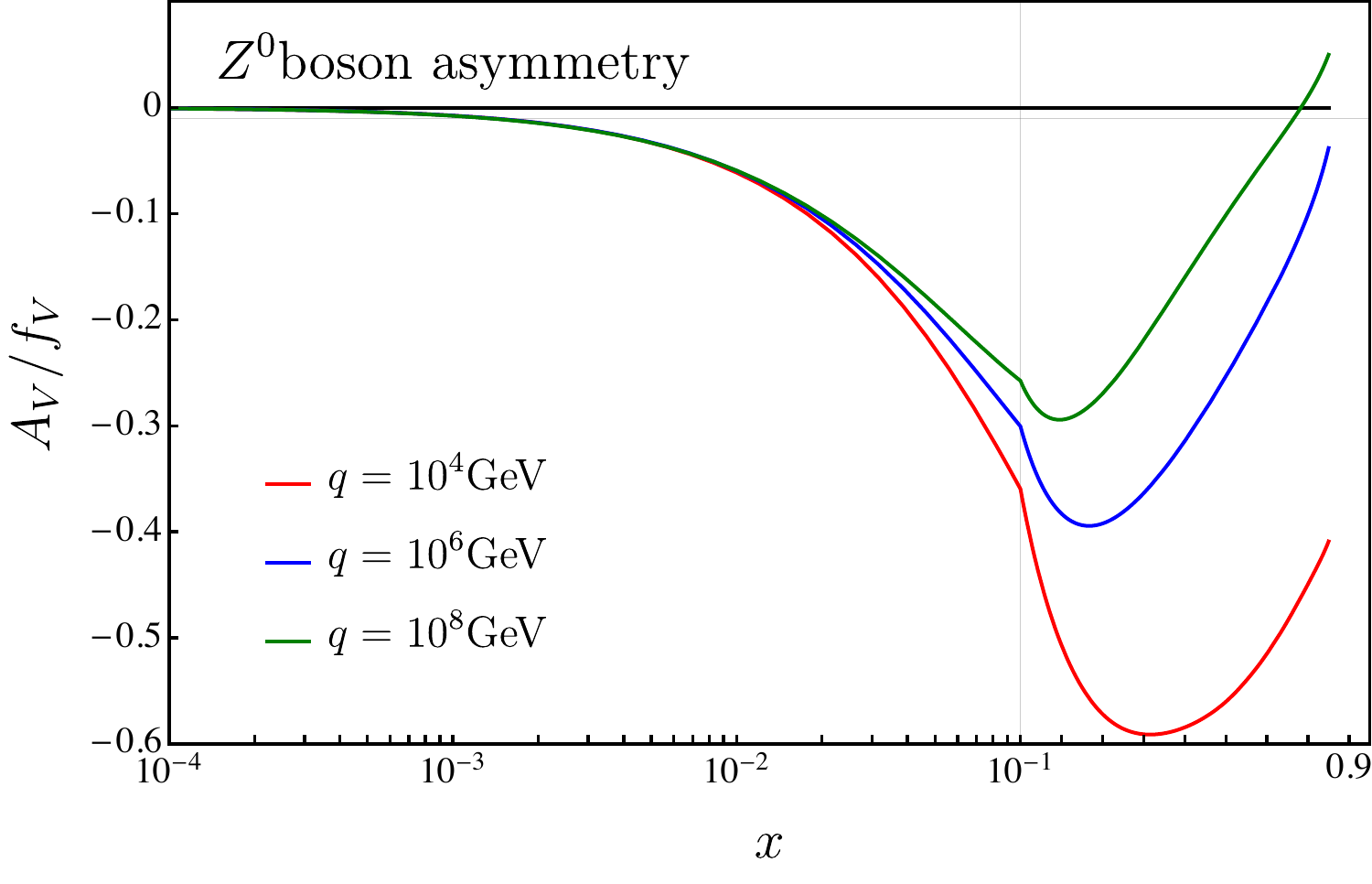}
	\includegraphics[scale=0.45]{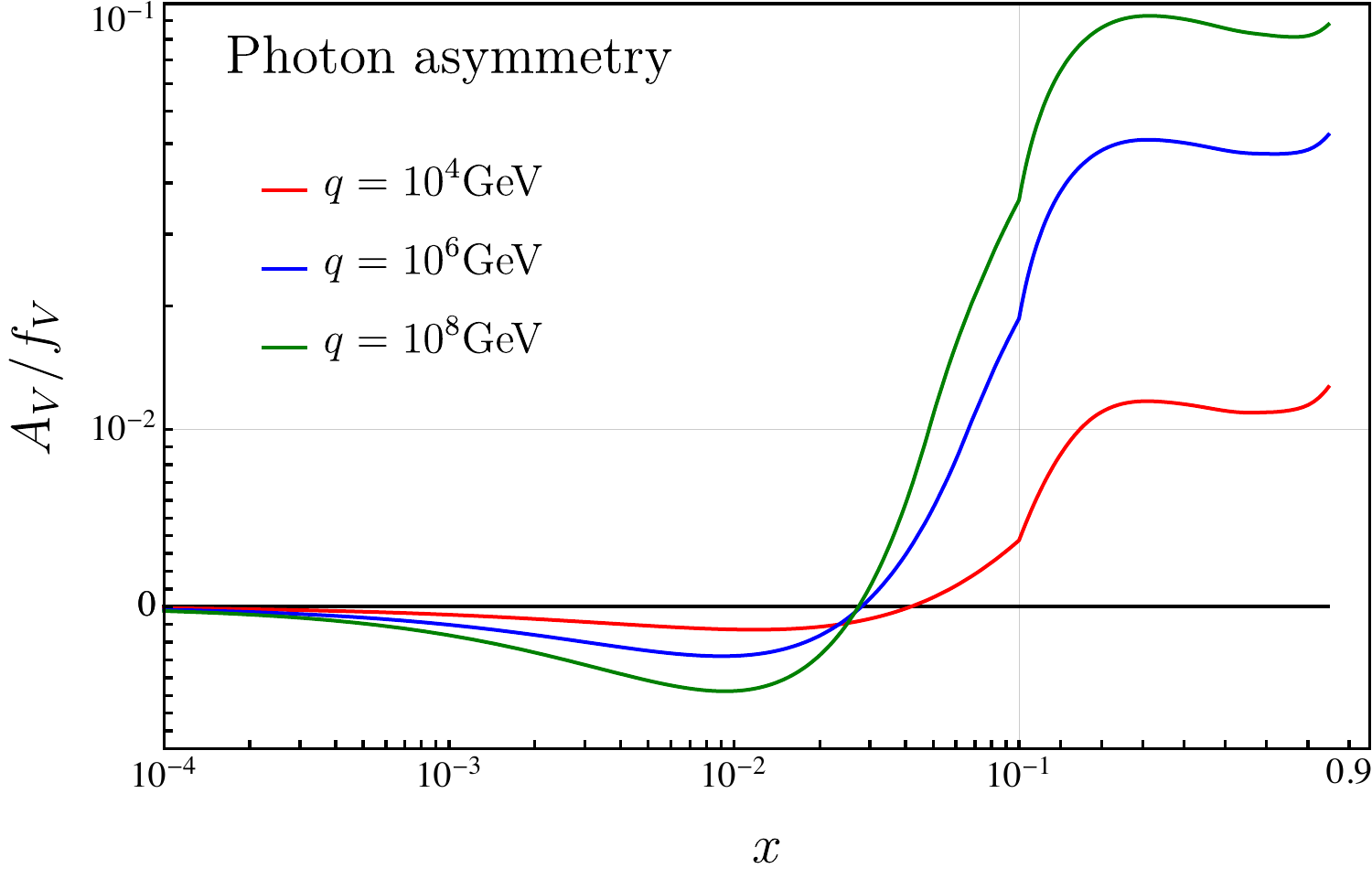}
	\includegraphics[scale=0.45]{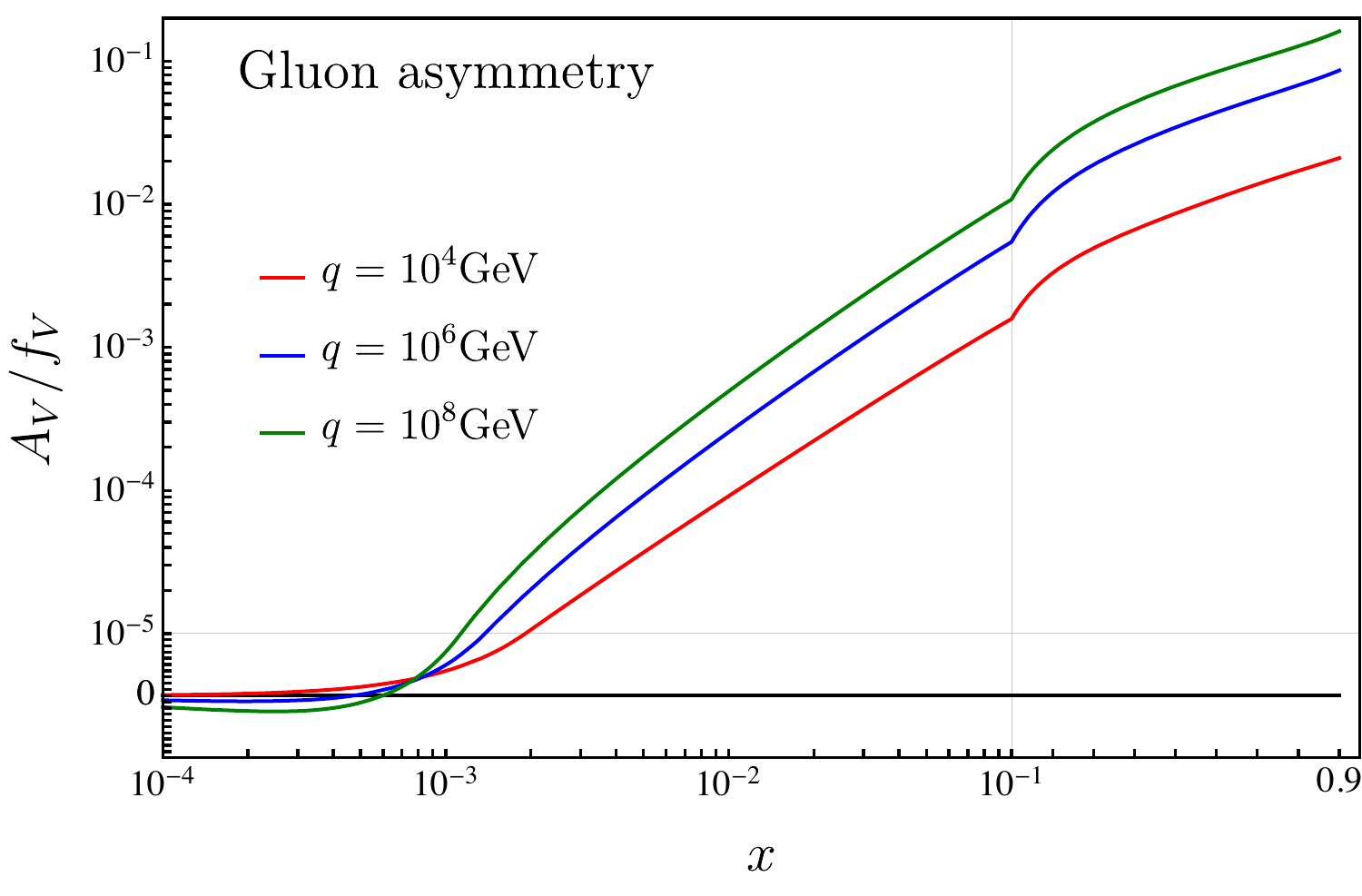}
	\caption{\label{fig:Polarization}%
		 Polarization of gauge bosons normalized to their
                 unpolarized PDFs. The thin gray line shows where the scales on the 
x- and/or y-axes switch between linear and logarithmic.
	}
}

In~\cite{Bauer:2017bnh} we presented results of the expansion of all PDFs, defining
\begin{align}
\label{eq:gdef}
\left[f_i^{\rm SM}(x,q)\right]_\alpha = f_i^{\rm noEW}(x,q) + g_i(x,q)
\,,
\end{align}
where 
\begin{align}
\label{eq:fNoEWDef}
f_i^{\rm noEW}(x,q) = \left\{ \begin{array}{l} \mbox{QCD+QED
                                evolution for $q < q_V$}\,, \\
                                \mbox{QCD evolution for $q > q_V$}\,. \end{array} \right.
\end{align}
and $\left[f_i^{\rm SM}(x,q)\right]_\alpha$ only includes the linear terms in $\alpha_{I\ne 3}$. 
These results were
used to match the resummed calculation to fixed-order results, and to understand the importance of the resummation and higher-order corrections that are very difficult to obtain in a fixed-order calculation. 
We have repeated the calculation of the first-order expansion of all PDFs, including all improvements discussed in this paper. While the numerical results change slightly, qualitatively all 
conclusions made in the previous paper remain unchanged. For this reason, we do not repeat the analysis here. We will, however, study the perturbative convergence of the parton luminosities, discussed next.

\FIGURE{
	\centering
	\includegraphics[scale=0.35]{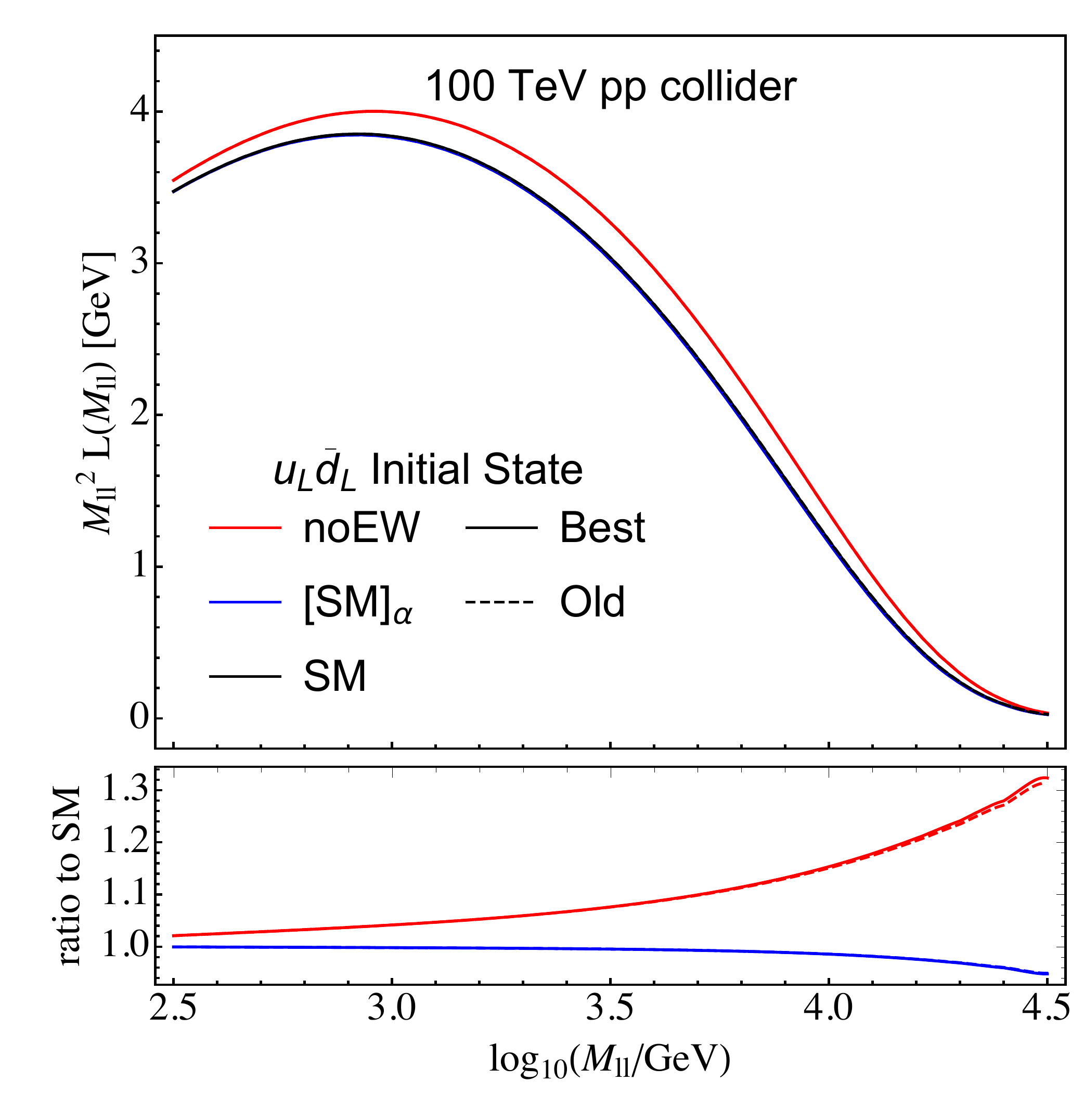}
	\includegraphics[scale=0.35]{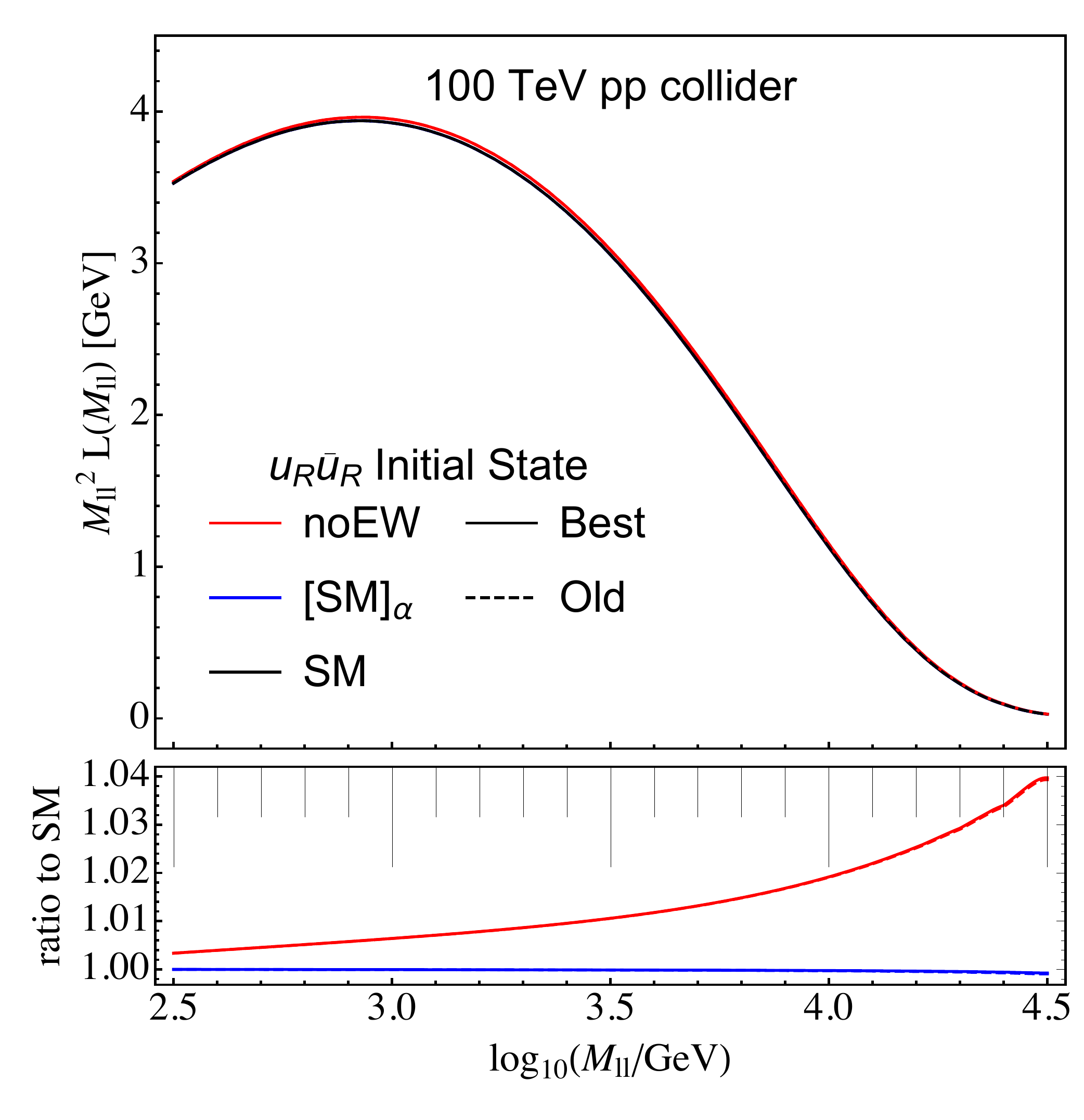}
	\includegraphics[scale=0.35]{Figures/LumiConvergence_uLdLbar_100.pdf}
	\includegraphics[scale=0.35]{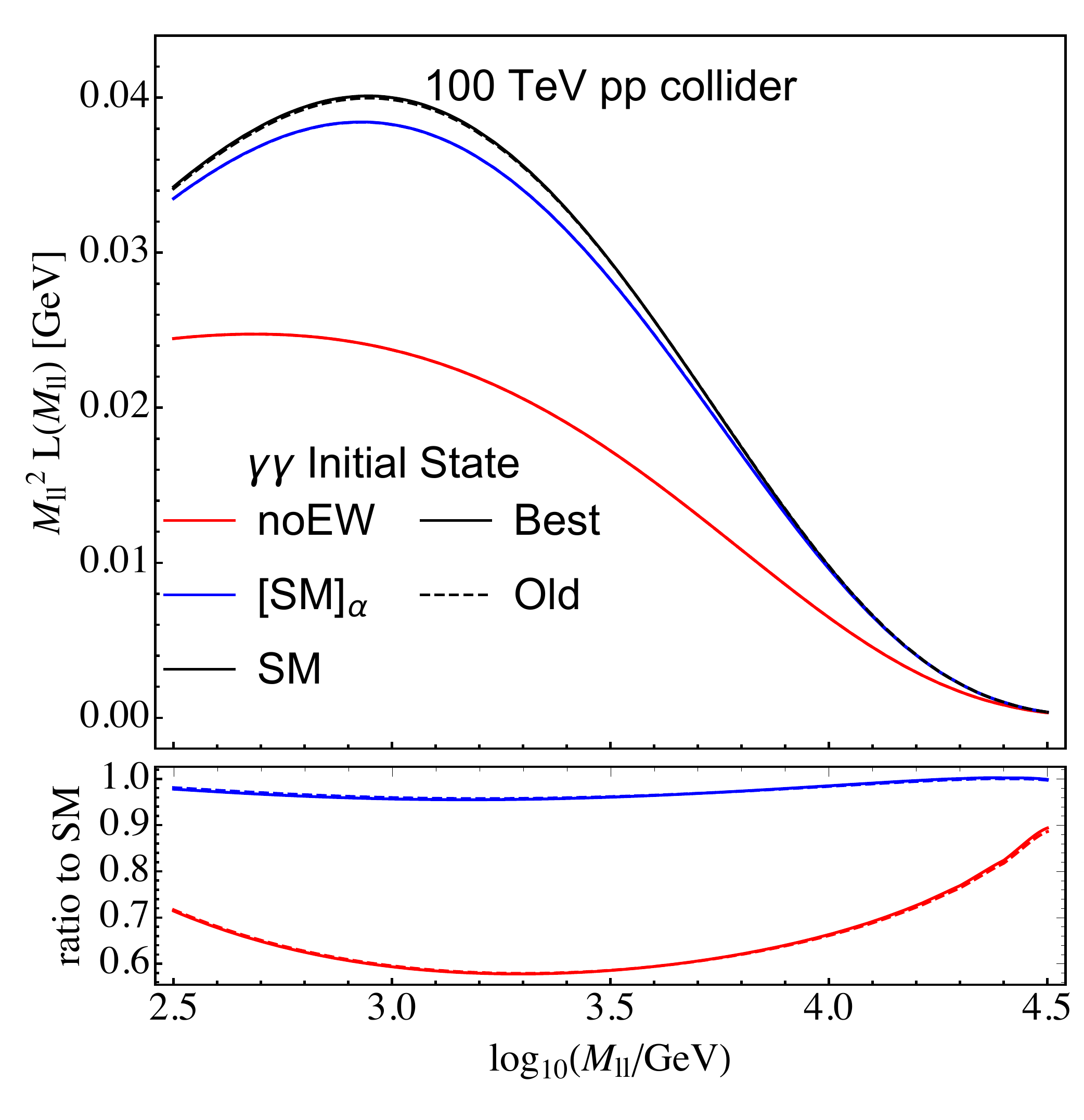}
	\includegraphics[scale=0.35]{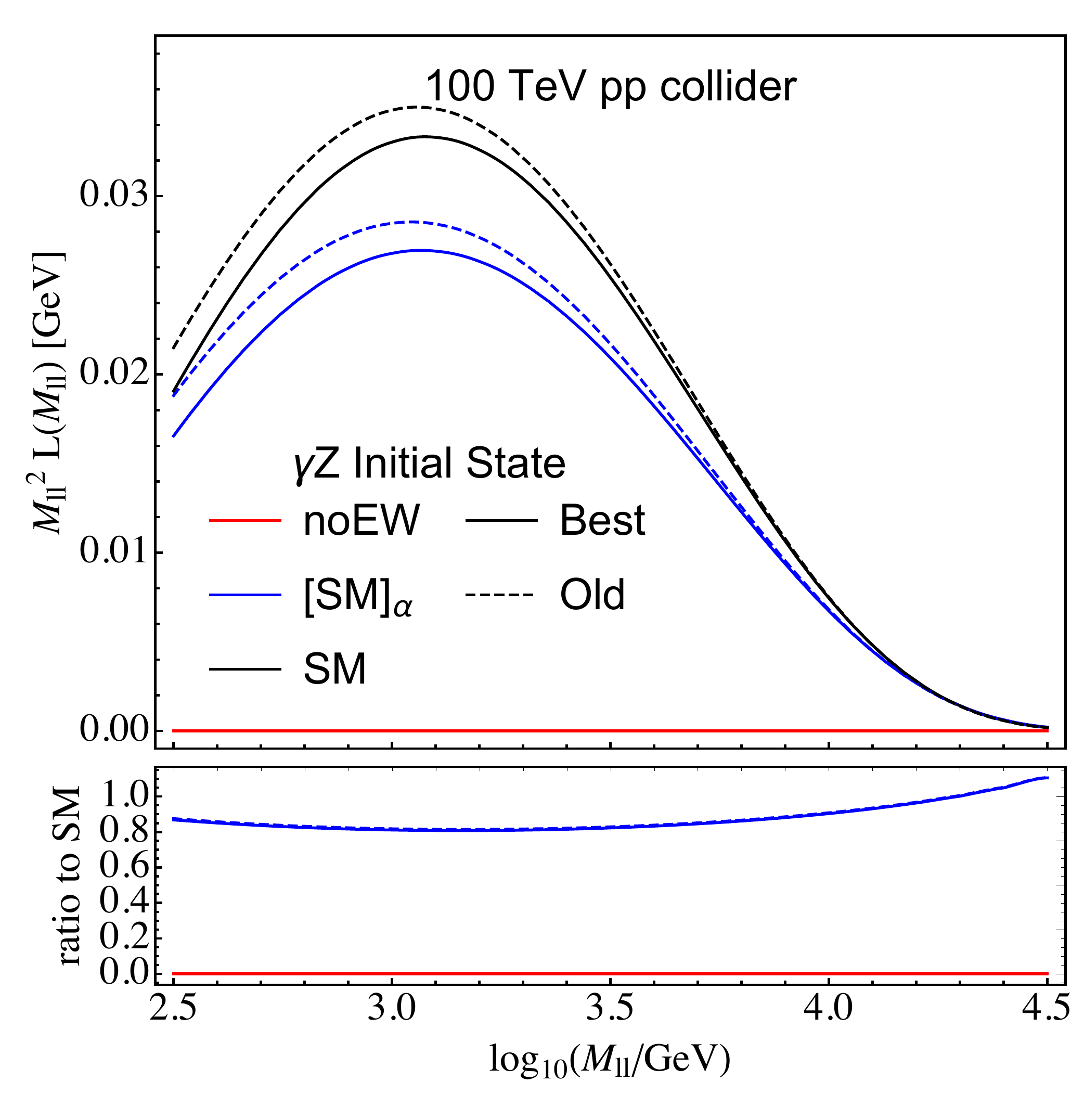}
	\includegraphics[scale=0.35]{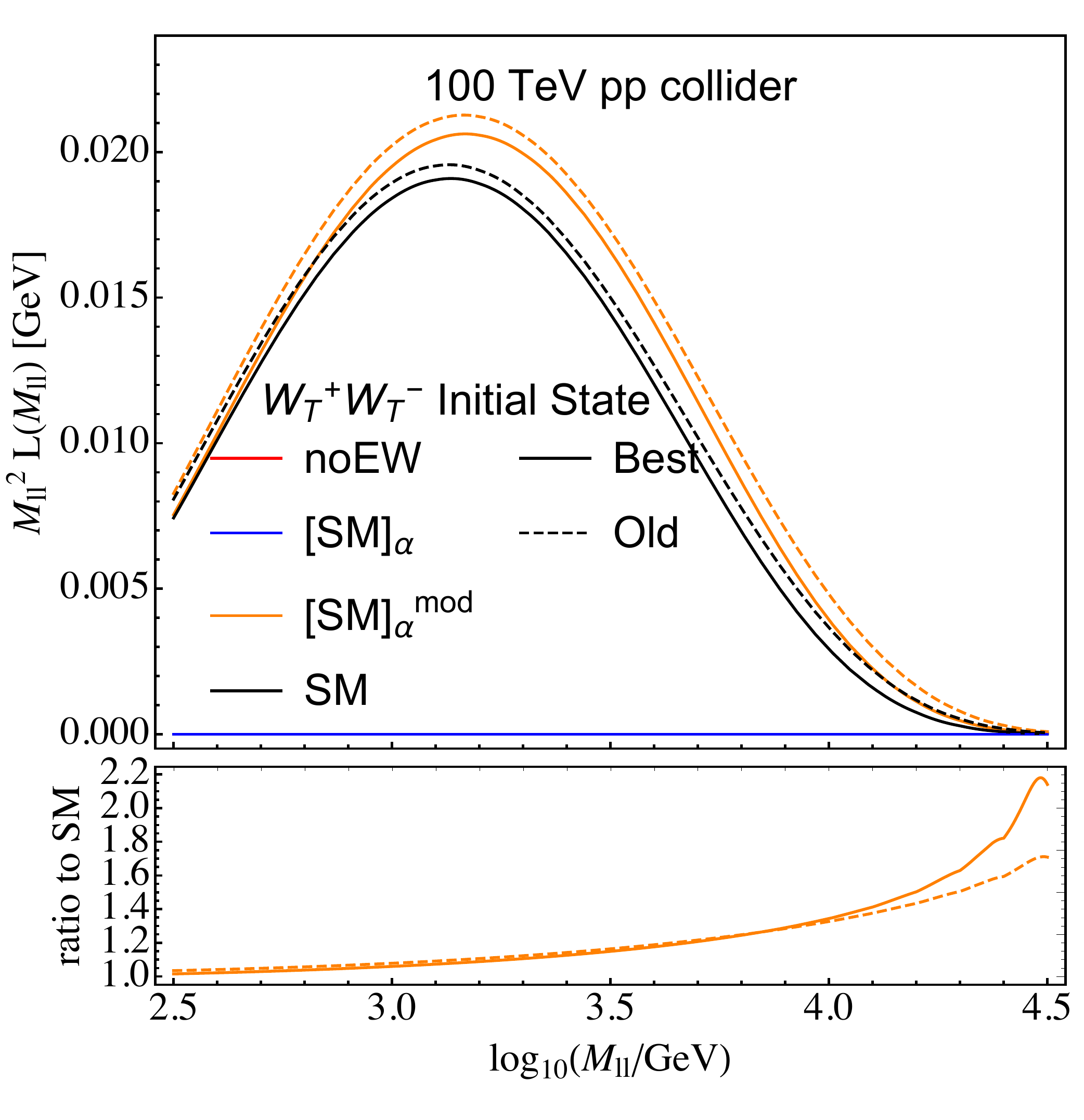}
	\caption{\label{fig:lumi_100}%
Plots showing luminosities for various choices of initial states. We show in black the luminosity computed using the full SM, in red the result without any EW effects, in blue the first order expansion and for $V_TV_T$ initial states in orange the luminosity when both first order expansions are multiplied together.}
}
As a final result, we combine the obtained PDFs into parton luminosities at a future 100 TeV $pp$ collider. In Fig.~\ref{fig:lumi_100} we show the results for a few selected parton luminosities
\begin{align}
\label{eq:Lumi_SM}
{\cal L}^{\rm SM}_{AB}(M_{\ell\ell}) = \int\! \df x_A\,\df x_B \, {\cal L}^{\rm SM}_{AB}\left(x_A, x_B; M_{\ell\ell}\right) \,\delta\left(M_{\ell\ell}-\sqrt{x_1x_2 S}\right)
\,,
\end{align}
with
\begin{align}
{\cal L}^{\rm SM}_{AB}(x_A, x_B; Q) &= f_A^{\rm SM}(x_A, Q) \, f_B^{\rm SM}(x_B, Q)
\,,
\end{align}
for $pp$ collisions at $\sqrt S=100$ TeV,
rescaled by the square of the invariant mass $M_{\ell\ell}$ to
overcome the steeply falling nature of the functions. 

For the transverse vector boson luminosities, one needs to consider
the positive and negative helicity PDFs of the bosons, such that there are
in general four different luminosities for each flavor combination. For
the production of fermions (after integrating over the rapidity of the
produced fermions), the relevant luminosity is the sum of $V_+ V_-$
and $V_- V_+$, which is related to the difference of the unpolarized and polarized luminosities
\beq
{\cal L}_{V V} - {\cal L}_{A_V A_V} = 2\left({\cal L}_{V_+ V_-} + {\cal L}_{V_- V_+}\right).
\eeq 
For this reason, we show this difference, but one has to remember that in general three more luminosities are required.  

For each figure, we show in black
${\cal L}^{\rm SM}$ (see \eq{Lumi_SM}). In red we show ${\cal L}^{\rm noEW}$, 
computed using PDFs that were evolved using only QCD and QED
interactions, as specified in Eq.~(\ref{eq:fNoEWDef}). In blue we show the 
$\left[{\cal L}^{\rm SM}\right]_\alpha$, given by
\begin{align}
\label{eq:LumiSMExpanded}
\left[ {\cal L}^{\rm SM}_{AB}(x_A, x_B; Q)\right]_\alpha &= f_A^{\rm noEW}(x_A, Q) \, f_B^{\rm noEW}(x_B, Q) + f_A^{\rm noEW}(x_A, Q) \, g_B(x_B, Q)
\nn
 & \quad + g_A(x_A, Q) \, f_B^{\rm noEW}(x_B, Q) 
\,,
\end{align}
and for VV initial states in
orange $\left[{\cal L}^{\rm SM}\right]_\alpha^{\rm mod}$, given by
\begin{align}
\label{eq:LumiSMExpandedVfusions}
\left[ {\cal L}^{\rm SM}_{AB}(x_A, x_B; Q)\right]_{\alpha}^{\rm mod} &= f_A^{\rm noEW}(x_A, Q) \, f_B^{\rm noEW}(x_B, Q) + f_A^{\rm noEW}(x_A, Q) \, g_B(x_B, Q)
\\
 & \quad + g_A(x_A, Q) \, f_B^{\rm noEW}(x_B, Q)  + g_A(x_A, Q) \, g_B(x_B, Q) \delta_{AB,V_TV_T}
 \notag
\,,\end{align}
which coincides with $\left[ {\cal L}^{\rm SM}_{AB}\right]_\alpha$ for all channels except $V_TV_T$. 

As for the PDFs, we show in solid lines the results including all effects discussed in this paper, and in dashed lines the results of~\cite{Bauer:2017bnh} that does not include these effects. 
For the $q \bar q$ and $\gamma\gamma$ luminosities the effects are so
small that two lines are practically indistinguishable. For
luminosities involving heavy vector boson PDFs, the effects are
larger, as can be expected from the results discussed for those PDFs
above. However, qualitatively, all conclusions
of~\cite{Bauer:2017bnh}, in particular about the importance of
resummation, are unchanged.

\section{Conclusions}
\label{sec:conc}

We have updated the results of
Refs.~\cite{Bauer:2017isx,Bauer:2017bnh} on parton distribution functions in the full SM by including 
three effects not considered in that earlier work. The first is the inclusion of gauge boson polarization, the second is to use non-zero input electroweak boson PDFs at the electroweak scale and the final effect is the improvement of the collinear evolution to full 
next-to-leading-order accuracy. 

Gauge boson polarizations arise because left- and right-handed fermions, which evolve differently in the full SM due to their 
different interactions with the SU(2) and U(1) gauge groups, couple differently to left-and right-handed polarized transverse 
vector bosons. This effect was first discussed in~\cite{Manohar:2018kfx}, where it was mentioned that it induces a polarization 
asymmetry in all transversely polarized gauge bosons. The implementation presented in this work shows that PDFs for the 
polarized $W_T$ and $Z_T$ bosons can be as large as their unpolarized PDFs, in particular at large $x$. 

In~\cite{Bauer:2017isx,Bauer:2017bnh} the initial conditions for the SM evolution were determined by treating the PDFs of quarks, gluons 
and the photon as non-zero at scale 10 GeV and then evolving them to
scale $q_0 \sim 100$ GeV using QCD and QED interactions. 
This meant that the PDFs for neutrinos, $W$ and $Z$ and Higgs bosons as well as the top quark were zero at $q_0$ and 
therefore only generated dynamically through the SM evolution. In this work, we take the results of~\cite{Fornal:2018znf} 
to obtain input values for the $W$ and $Z$ bosons (both longitudinal and transverse) at $q_0$. This therefore combines 
the resummation of the large logarithmic terms generated by the evolution with the threshold effects obtained from the 
fixed order results at $q_0$. As shown, this changes the results for electroweak vector bosons at low values of $q$, but 
these effects become subdominant at large values of $q$.

The final effect is the improvement of the collinear evolution to full next-to-leading-order accuracy. This was already 
discussed for fragmentation functions in~\cite{Bauer:2018xag}, and can be implemented through a proper definition 
of the running coupling constant. Such higher logarithmic resummation becomes most important at scales for 
which $\alpha L \sim 1$, which requires extremely large values of $q \sim 10^{15}$ GeV. Thus, one expects that 
the higher logarithmic effects give rise to only small effects at phenomenologically relevant scales, which is 
confirmed by our implementation. 

\acknowledgments
We thank Aneesh Manohar and Wouter Waalewijn for valuable discussions.
This work was supported by the Director, Office of Science, Office of
High Energy Physics of the U.S. Department of Energy under the
Contract No. DE-AC02-05CH11231 (CWB), and partially supported by
STFC consolidated grants ST/L000385/1 and ST/P000681/1 (BRW).

\addcontentsline{toc}{section}{References}
\bibliographystyle{JHEP}
\bibliography{../SMevol_paper}

\providecommand{\href}[2]{#2}\begingroup\raggedright\begin{thebibliography}{10}

\bibitem{Bauer:2017isx}
C.~W. Bauer, N.~Ferland and B.~R. Webber, \emph{{Standard Model Parton
  Distributions at Very High Energies}},
  \href{http://dx.doi.org/10.1007/JHEP08(2017)036}{\emph{JHEP} {\bf 08} (2017)
  036}, [\href{http://arxiv.org/abs/1703.08562}{{\tt 1703.08562}}].

\bibitem{Bauer:2017bnh}
C.~W. Bauer, N.~Ferland and B.~R. Webber, \emph{{Combining initial-state
  resummation with fixed-order calculations of electroweak corrections}},
  \href{http://dx.doi.org/10.1007/JHEP04(2018)125}{\emph{JHEP} {\bf 04} (2018)
  125}, [\href{http://arxiv.org/abs/1712.07147}{{\tt 1712.07147}}].

\bibitem{Bauer:2018xag}
C.~W. Bauer, D.~Provasoli and B.~R. Webber, \emph{{Standard Model Fragmentation
  Functions at Very High Energies}},
  \href{http://arxiv.org/abs/1806.10157}{{\tt 1806.10157}}.

\bibitem{Manohar:2018kfx}
A.~V. Manohar and W.~J. Waalewijn, \emph{{Electroweak Logarithms in Inclusive
  Cross Sections}},  \href{http://arxiv.org/abs/1802.08687}{{\tt 1802.08687}}.

\bibitem{Fornal:2018znf}
B.~Fornal, A.~V. Manohar and W.~J. Waalewijn, \emph{{Electroweak Gauge Boson
  Parton Distribution Functions}},  \href{http://arxiv.org/abs/1803.06347}{{\tt
  1803.06347}}.

\bibitem{Manohar:2016nzj}
A.~Manohar, P.~Nason, G.~P. Salam and G.~Zanderighi, \emph{{How bright is the
  proton? A precise determination of the photon parton distribution function}},
  \href{http://dx.doi.org/10.1103/PhysRevLett.117.242002}{\emph{Phys. Rev.
  Lett.} {\bf 117} (2016) 242002}, [\href{http://arxiv.org/abs/1607.04266}{{\tt
  1607.04266}}].

\bibitem{Manohar:2017eqh}
A.~V. Manohar, P.~Nason, G.~P. Salam and G.~Zanderighi, \emph{{The Photon
  Content of the Proton}},
  \href{http://dx.doi.org/10.1007/JHEP12(2017)046}{\emph{JHEP} {\bf 12} (2017)
  046}, [\href{http://arxiv.org/abs/1708.01256}{{\tt 1708.01256}}].

\bibitem{Ciafaloni:2005fm}
P.~Ciafaloni and D.~Comelli, \emph{{Electroweak evolution equations}},
  \href{http://dx.doi.org/10.1088/1126-6708/2005/11/022}{\emph{JHEP} {\bf 11}
  (2005) 022}, [\href{http://arxiv.org/abs/hep-ph/0505047}{{\tt
  hep-ph/0505047}}].

\bibitem{Dokshitzer:1978hw}
Y.~L. Dokshitzer, D.~Diakonov and S.~I. Troian, \emph{{Hard Processes in
  Quantum Chromodynamics}},
  \href{http://dx.doi.org/10.1016/0370-1573(80)90043-5}{\emph{Phys. Rept.} {\bf
  58} (1980) 269--395}.

\bibitem{Amati:1980ch}
D.~Amati, A.~Bassetto, M.~Ciafaloni, G.~Marchesini and G.~Veneziano, \emph{{A
  Treatment of Hard Processes Sensitive to the Infrared Structure of QCD}},
  \href{http://dx.doi.org/10.1016/0550-3213(80)90012-7}{\emph{Nucl. Phys.} {\bf
  B173} (1980) 429--455}.

\bibitem{Catani:1990rr}
S.~Catani, B.~R. Webber and G.~Marchesini, \emph{{QCD coherent branching and
  semiinclusive processes at large x}},
  \href{http://dx.doi.org/10.1016/0550-3213(91)90390-J}{\emph{Nucl. Phys.} {\bf
  B349} (1991) 635--654}.

\bibitem{Chiu:2007dg}
J.-y. Chiu, F.~Golf, R.~Kelley and A.~V. Manohar, \emph{{Electroweak
  Corrections in High Energy Processes using Effective Field Theory}},
  \href{http://dx.doi.org/10.1103/PhysRevD.77.053004}{\emph{Phys. Rev.} {\bf
  D77} (2008) 053004}, [\href{http://arxiv.org/abs/0712.0396}{{\tt
  0712.0396}}].

\bibitem{Schmidt:2015zda}
C.~Schmidt, J.~Pumplin, D.~Stump and C.~P. Yuan, \emph{{CT14QED parton
  distribution functions from isolated photon production in deep inelastic
  scattering}}, \href{http://dx.doi.org/10.1103/PhysRevD.93.114015}{\emph{Phys.
  Rev.} {\bf D93} (2016) 114015}, [\href{http://arxiv.org/abs/1509.02905}{{\tt
  1509.02905}}].

\bibitem{Butterworth:2015oua}
J.~Butterworth et~al., \emph{{PDF4LHC recommendations for LHC Run II}},
  \href{http://dx.doi.org/10.1088/0954-3899/43/2/023001}{\emph{J. Phys.} {\bf
  G43} (2016) 023001}, [\href{http://arxiv.org/abs/1510.03865}{{\tt
  1510.03865}}].

\bibitem{Chen:2016wkt}
J.~Chen, T.~Han and B.~Tweedie, \emph{{Electroweak Splitting Functions and High
  Energy Showering}},
  \href{http://dx.doi.org/10.1007/JHEP11(2017)093}{\emph{JHEP} {\bf 11} (2017)
  093}, [\href{http://arxiv.org/abs/1611.00788}{{\tt 1611.00788}}].

\end{thebibliography}\endgroup

\end{document}